%% file: template.tex
\documentclass[10pt,journal,compsoc]{IEEEtran}
\ifCLASSOPTIONcompsoc
  \usepackage[nocompress]{cite}
\else
  \usepackage{cite}
\fi
\ifCLASSINFOpdf
\else
\fi

\usepackage{graphicx}                
\DeclareGraphicsExtensions{.pdf,.png,.jpg,.jpeg,.eps} 

\graphicspath{{figures/}{pictures/}{images/}{./}} 

\usepackage{microtype}                 
\PassOptionsToPackage{warn}{textcomp}  
\usepackage{textcomp}                  
\usepackage{mathptmx}                  
\usepackage{times}                     
\usepackage{cite}                      
\usepackage{tabu}                      
\usepackage{booktabs}                  
\usepackage{xcolor}
\usepackage{amsmath}
\usepackage{amssymb}
\usepackage{multirow}
\usepackage{multicol}
\usepackage{booktabs}  
\usepackage{threeparttable}  
\usepackage{tikz}

\usepackage{url}
\usepackage{enumitem}
\usepackage{paralist}                  

\usepackage{setspace} 
\usepackage{hyperref}
\hypersetup{hidelinks}

\newcommand{\name}{{\scshape Somnus}} 

\newcommand{\etal}{et~al.\ }
\newcommand{\eg}{e.g.}
\newcommand{\ie}{i.e.}

\definecolor{cText}{HTML}{45bce7}
\definecolor{cVis}{HTML}{ed7d31}
\definecolor{cAnswer}{HTML}{C1E8C5}
\definecolor{kLime}{HTML}{02b340}
\newcommand{\kai}[1]{\textcolor{black}{#1}}

\newcommand{\revise}[1]{\textcolor{black}{#1}}

\usepackage{graphicx}
\newlength{\myMheight}
\begin{document}
%
\title{Visualizing the Scripts of Data Wrangling \\ with \name{}}
%
%
%
%

\author{Kai~Xiong, 
        Siwei~Fu, 
        Guoming~Ding, 
        Zhongsu~Luo, 
        Rong~Yu, 
        Wei~Chen, 
        Hujun~Bao, 
        Yingcai~Wu 

\IEEEcompsocitemizethanks{\IEEEcompsocthanksitem K. Xiong, G. Ding, W. Chen, H. Bao, and Y. Wu are with the State
Key Lab of CAD\&CG, Zhejiang University, Hangzhou, China, and with Zhejiang Lab, Hangzhou, China.
E-mails: \{kaixiong, dinggm, chenvis\}@zju.edu.cn, bao@cad.zju.edu.cn, ycwu@zju.edu.cn.
\IEEEcompsocthanksitem S. Fu and R. Yu are with the Zhejiang Lab, Hangzhou, China.
\protect\\
E-mail: fusiwei339@gmail.com, 1721298964@qq.com.

\IEEEcompsocthanksitem Z. Luo is with Zhejiang University of Technology, Hangzhou, China, and also with the Zhejiang Lab, Hangzhou, China.
\protect\\
E-mail: rickyluozs@gmail.com.

\IEEEcompsocthanksitem Yingcai Wu and Siwei Fu are the co-corresponding authors.

}
\thanks{Manuscript received xx xxx, 20xx; revised xx xxx, 20xx.}
}

%
%

\markboth{Journal of \LaTeX\ Class Files,~Vol.~14, No.~8, August~2015}%
{Shell \MakeLowercase{\textit{et al.}}: Bare Demo of IEEEtran.cls for Computer Society Journals}
\IEEEtitleabstractindextext{%
\begin{abstract}
Data workers use various scripting languages for data transformation, such as SAS, R, and Python. 
However, understanding intricate code pieces requires advanced programming skills, which hinders data workers from grasping the idea of data transformation at ease.
Program visualization is beneficial for debugging and education and has the potential to illustrate transformations intuitively and interactively. 
In this paper, we explore visualization design for demonstrating the semantics of code pieces in the context of data transformation. 
First, to depict individual data transformations, we structure a design space by two primary dimensions, \ie, key parameters to encode and possible visual channels to be mapped. 
Then, we derive a collection of $23$ glyphs that visualize the semantics of transformations.
Next, we design a pipeline, named \name{}, that provides an overview of the creation and evolution of data tables using a provenance graph.
At the same time, it allows detailed investigation of individual transformations.
User feedback on \name{} is positive. 
Our study participants achieved better accuracy with less time using \name{}, and preferred it over \revise{carefully-crafted} textual description.
Further, we provide two example applications to demonstrate the utility and versatility of \name{}.
\end{abstract}

\begin{IEEEkeywords}
Program understanding, data transformation, visualization design.
\end{IEEEkeywords}}

\maketitle

\IEEEdisplaynontitleabstractindextext

%
\IEEEpeerreviewmaketitle

\IEEEraisesectionheading{\section{Introduction}\label{sec:introduction}}

\input{src/intro}
\input{src/relatedwork}

\input{src/requirement}
\input{src/glyph}

\input{src/pipeline_design}
\input{src/evaluation}
\input{src/discussion}

\section*{Acknowledgments}
This work was supported by NSFC (62072400, 62002331) and the Collaborative Innovation Center of Artificial Intelligence by MOE and Zhejiang Provincial Government (ZJU).
The work was also partially funded by the Zhejiang Lab (2021KE0AC02, 2020KE0AA02).
We are grateful to our study participants and anonymous reviewers for their insightful feedback.


\ifCLASSOPTIONcaptionsoff
  \newpage
\fi



%



\bibliographystyle{abbrv-doi}
\bibliography{template}

%

\begin{IEEEbiography}[{\includegraphics[width=1in,height=1.25in,clip,keepaspectratio]{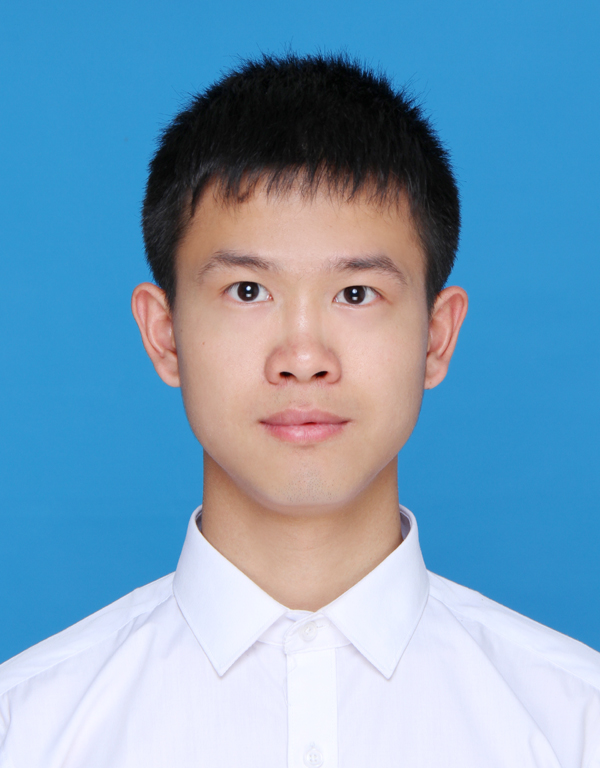}}]{Kai Xiong}
  is a Ph.D. student at the State Key Laboratory of CAD\&CG, Zhejiang University, and works under the supervision of Prof. Yingcai Wu. He holds a bachelor's degree in Computer Science from Xidian University. His main research interests center on visual analytics and data wrangling. He is also interested in how to apply artificial intelligence to data visualization.
\end{IEEEbiography}


\begin{IEEEbiography}[{\includegraphics[width=1in,height=1.25in,clip,keepaspectratio]{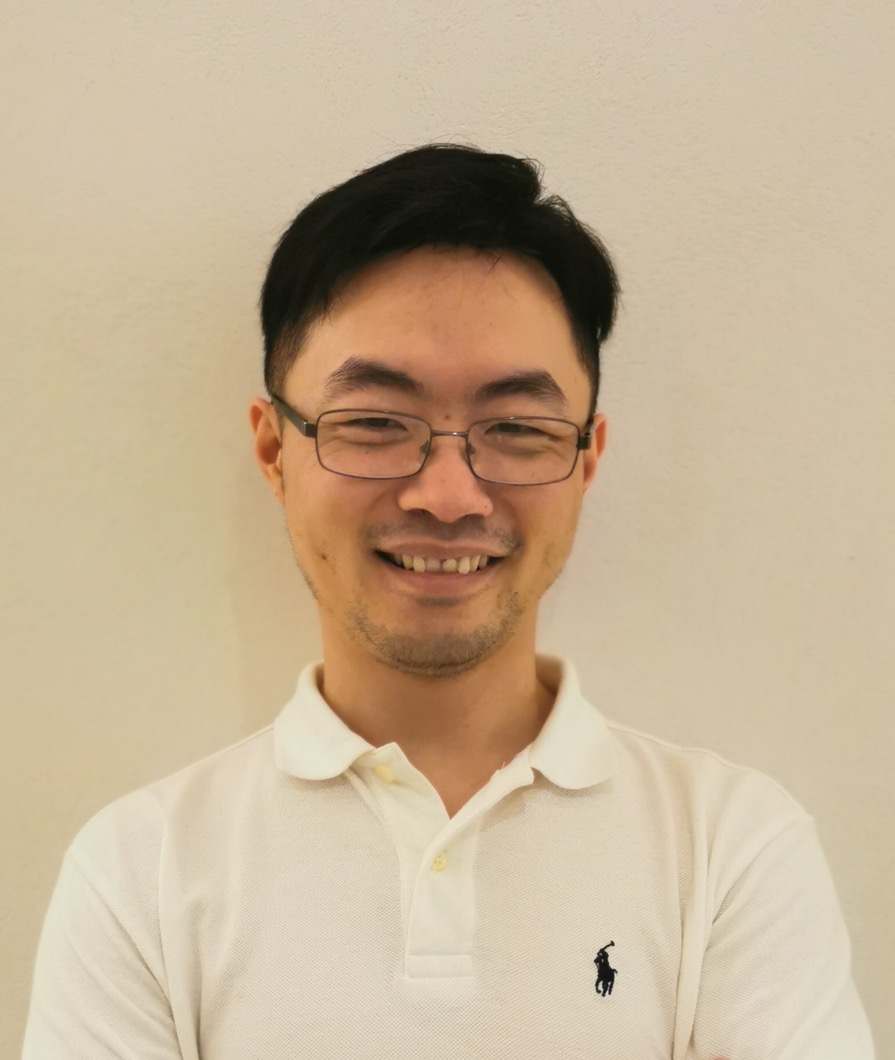}}]{Siwei Fu}
 is an associate research scientist in Zhejiang Lab. His main research interests include: visual analytics, intelligent user interface, and natual language interface. He received his Ph.D. degree in Computer Science and Engineering from the Hong Kong University of Science and Technology. For more information, please visit https://fusiwei339.bitbucket.io/
\end{IEEEbiography}


\begin{IEEEbiography}[{\includegraphics[width=1in,height=1.25in,clip,keepaspectratio]{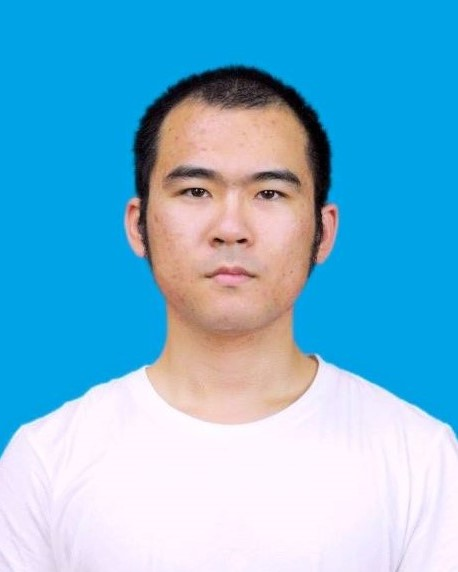}}]{Guoming Ding}
  received his B.S. degree in Mechanical Engineering from Xi’an Jiaotong University in 2020. He is currently pursuing the master's degree with the State Key Lab of CAD\&CG, Zhejiang University. His research interests mainly include the visualization and causal analysis.
 \end{IEEEbiography}


 \begin{IEEEbiography}[{\includegraphics[width=1in,height=1.25in,clip,keepaspectratio]{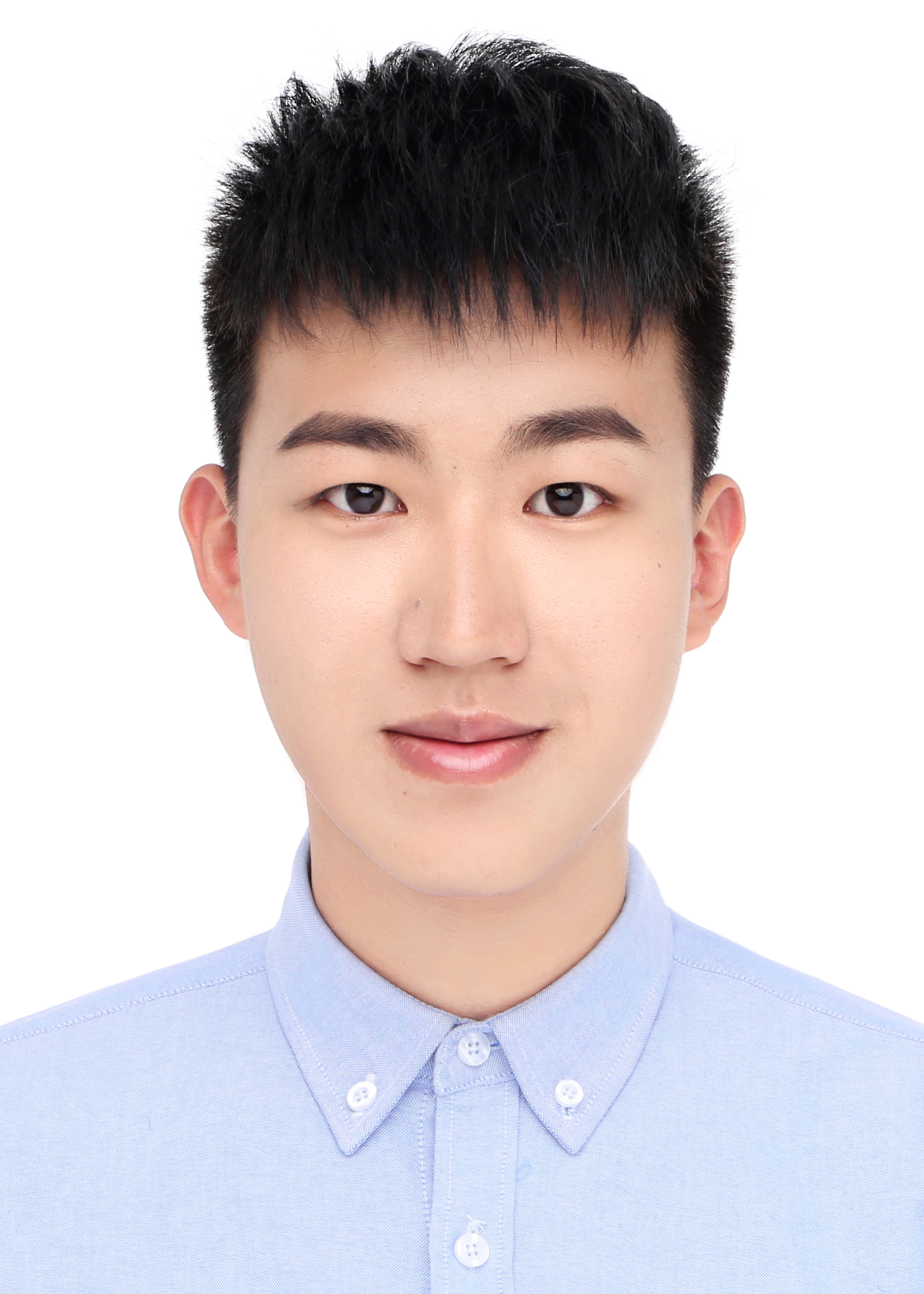}}]{Zhongsu Luo}
  received his B.S. degree in Software Engineering from Zhejiang University of Technology in 2020. He is currently pursuing the master’s degree in Zhejiang University of Technology. His research interests include the visualization, and visual analysis.
 \end{IEEEbiography}


 \begin{IEEEbiography}[{\includegraphics[width=1in,height=1.25in,clip,keepaspectratio]{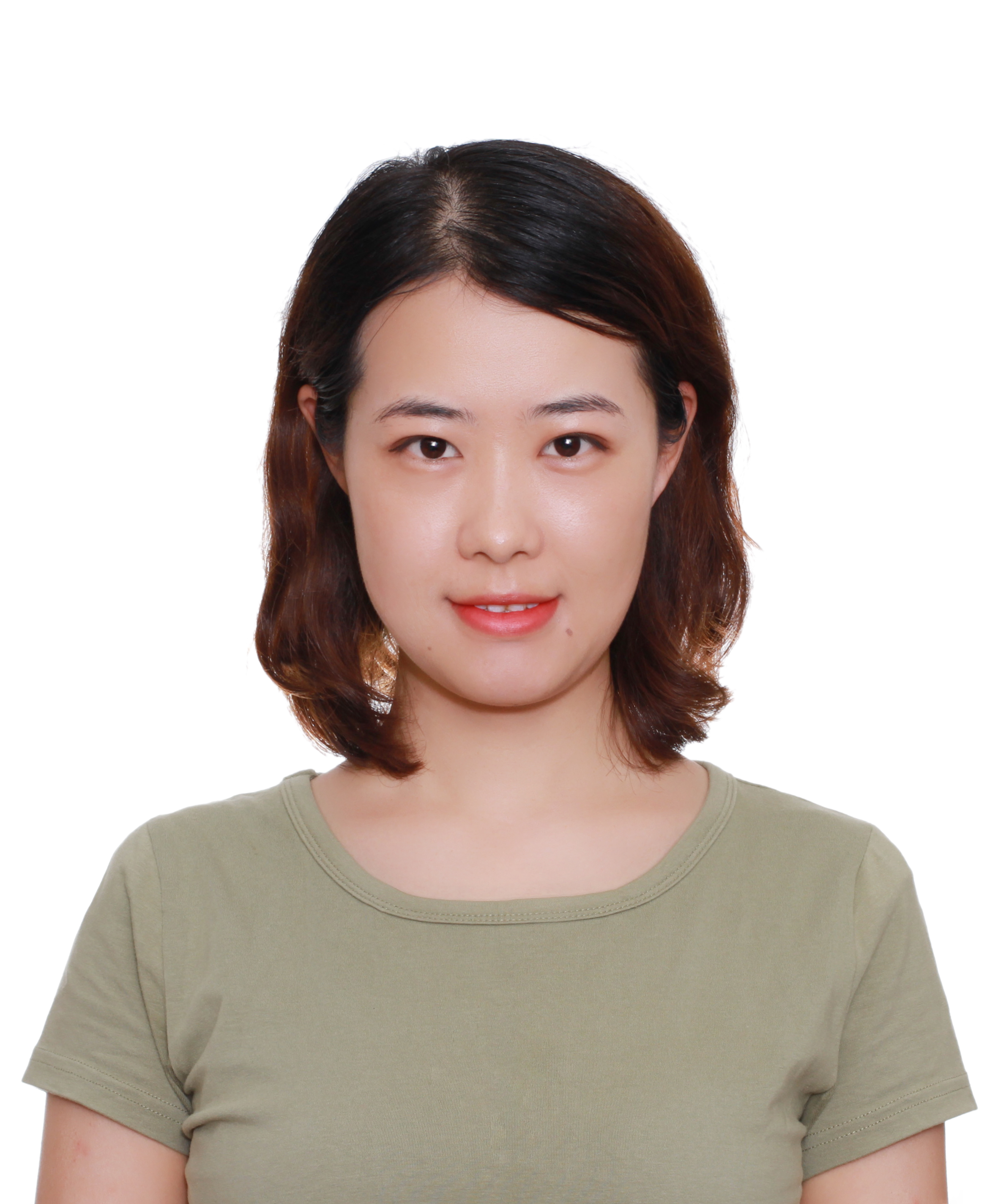}}]{Rong Yu} 
  received the master's degree in visual communication design from Hangzhou Normal University, China, in 2015. She is currently a senior research engineer in Zhejiang Lab, China. She has eight years of expertise in graphic design and UI/UX design. For more information, please visit https://dribbble.com/yurongt
 \end{IEEEbiography}


 \begin{IEEEbiography}[{\includegraphics[width=1in,height=1.25in,clip,keepaspectratio]{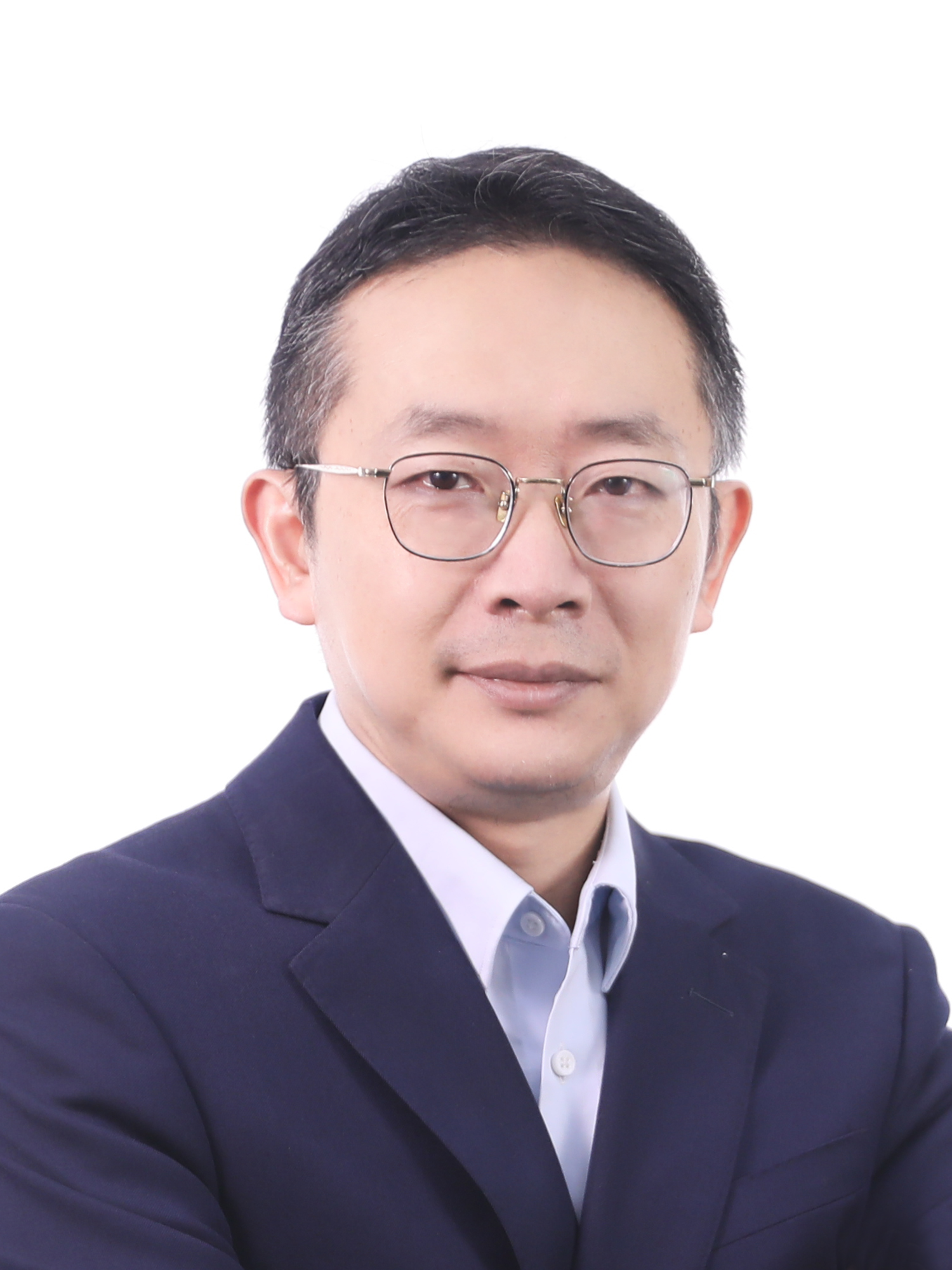}}]{Wei Chen}
  is a professor at the State Key Lab of CAD\&CG, Zhejiang University. His research interests are visualization and visual analysis. He has published more than 30 IEEE/ACM Transactions and IEEE VIS papers. He actively served as a guest or associate editor of IEEE Transactions on Visualization and Computer Graphics, IEEE Transactions on IntelligentTransportation Systems, and Journal of Visualization. For more information, please refer to http://www.cad.zju.edu.cn/home/chenwei/
 \end{IEEEbiography}

 
 \begin{IEEEbiography}[{\includegraphics[width=1in,height=1.25in,clip,keepaspectratio]{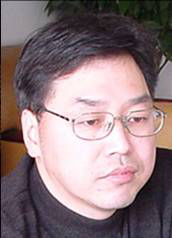}}]{Hujun Bao}
   is a professor with the State Key Laboratory of CAD\&CG and the College of Computer Science and Technology, Zhejiang University Zhejiang, China. He leads the 3D graphics computing group in the lab, which mainly makes researches on geometry computing, 3D visual computing, real-time rendering, and their applications. His research goal is to investigate the fundamental theories and algorithms to achieve good visual perception for interactive digital environments, and develop related systems.
 \end{IEEEbiography}


\begin{IEEEbiography}[{\includegraphics[width=1in,height=1.25in,clip,keepaspectratio]{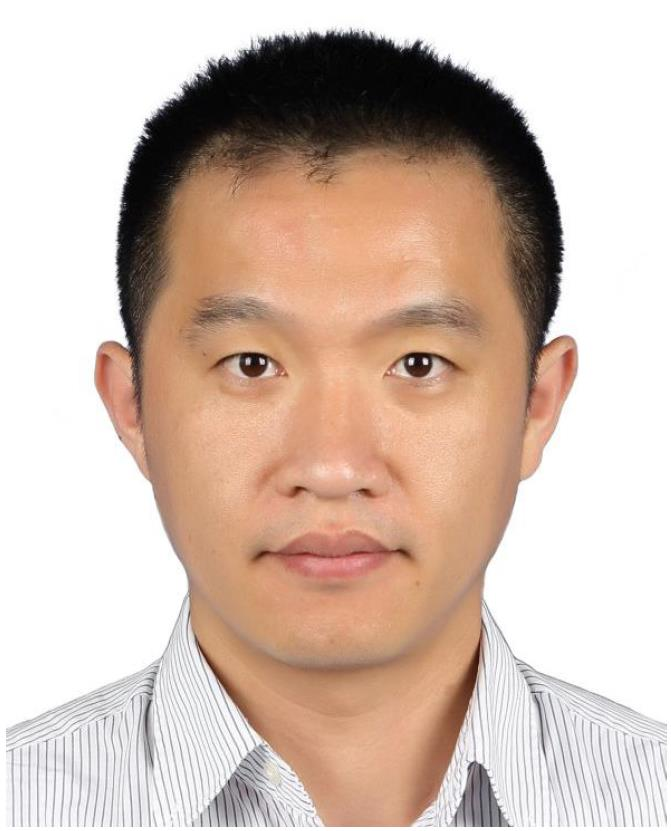}}]{Yingcai Wu} 
  is a professor at the State Key Lab of CAD\&CG, Zhejiang University, China. His primary research interests lie in information visualization and visual analytics, with focuses on sports science and urban computing. He received his Ph.D. degree in Computer Science from the Hong Kong University of Science and Technology. Prior to his current position, Dr. Wu was a postdoctoral researcher at the University of California, Davis from 2010 to 2012, and a researcher in Microsoft Research Asia from 2012 to 2015. For more information, please visit http://www.ycwu.org
 \end{IEEEbiography}





\vfill


\end{document}

%% file: src/intro.tex
\IEEEPARstart{S}{cripting} languages including SAS, R, and Python have been widely accepted by data workers for data transformation.
They usually seek to understand the semantics of scripts in various scenarios.
For example, validation (or called double-checking in some companies and laboratories) is important for data scientists.
A data scientist might seek to understand code pieces written by others, then locate and correct possible mistakes.
Understanding the semantics of an intricate script, however, requires advanced programming skills. 
And sometimes, the process is tedious and error-prone~\cite{lewis1987can, sorva2013review, qian2017students}.


A number of program visualization techniques have been proposed for debugging and communication. 
For example, some techniques, such as Whyline~\cite{ko2004designing}, Timelapse~\cite{burg2013interactive}, and FireCrystal~\cite{oney2009firecrystal}, utilize visualizations to help programmers identify and fix bugs. 
Those debugging tools focus on revealing the runtime behavior, such as the values of objects and variables, on allowing programmers to inspect the program state.
However, depicting program states benefits little in communicating the semantics of code pieces.
Others, such as algorithm visualizations~\cite{shaffer2007algorithm, Mike2014VA} and automatic generation of flowcharts~\cite{AutoFlowchart, code2flow, cheon2019vizme}, aim to help learners understand the flow of algorithms.
However, little attention has been paid to illustrating the process of data transformation.

In this work, we explore visualization design for depicting the semantics of code pieces in the context of data transformation.
To present individual data transformations, we first outline a design space consisting of two primary dimensions, \ie, key parameters to encode and potential visual channels that can be mapped. 
Then, we propose a collection of $23$ glyphs that demonstrates the semantics of transformations.
Given a code piece containing a series of functions, data tables are created and changed.
To illustrate the evolution of tables, we contribute the design and implementation of \name{},
a pipeline that accepts a script and data tables as input and results in a graph model where nodes are tables while edges are data transformations.
\name{} consists of two main components, \ie, Program Adaptor and Visualization Generator.
The Program Adaptor parses code pieces, generates input and output tables for each statement, and infer transformations based on rules.
On the other hand, the Visualization Generator creates visual representations to illustrate data provenance.
\revise{We claim that \name{} facilitates the understanding of intricate code pieces, including the semantics of individual operations and table dependencies of the entire data wrangling process.
To some extent, \name{} supports some higher-level tasks such as helping users debug programs of data wrangling and correct errors in the code.
Besides, the idea of \name{} is general and can be adapted to various scripting languages, including R, Python, etc.}

\revise{
To evaluate the effectiveness of the glyph design and \name{}, we conducted a controlled study to compare our visual representations with carefully-crafted textual descriptions.
The results show that our participants can understand complex wrangling scripts more accurately in a shorter time
and prefer our visualizations in terms of helpfulness and interpretability.}
In addition, we demonstrate the utility and versatility of \name{} with two example applications. 
The first application shows how \name{} can be adapted to Python and facilitates \revise{the validation of a piece of wrangling script}.
As for the second application, \name{} is adapted to R and used to support {\scshape Morpheus}~\cite{feng2017component} in interactive data transformation.

To conclude, the contributions in this paper include:
1) a design space consisting of two dimensions that guide the design of a collection of $23$ glyphs,
2) a pipeline, called \name{}, that visualizes the creation and evolution of data tables across a series of transformations,
3) a controlled study that evaluates how users perform with visualization and text using comparison tasks, and 
4) two example applications that showcase how \name{} can benefit different usage scenarios. 

%% file: src/relatedwork.tex
\section{Related Work}

\subsection{Program Visualization}
Program visualization refers to \textit{``the visualization of actual program code or data structures in either static or dynamic form''}~\cite{urquiza2004survey}.
\revise{
In program visualization, different audiences vary in analytical tasks, which require tailored visual representations~\cite{chotisarn2020systematic}.
For example, software engineers and data scientists are dedicated to development activities including programming, debugging, testing, etc.
Systems focusing on these activities usually need to visualize the runtime behavior of the program including object states, function calls, etc.
Another example is that, data workers~\cite{liu2019understanding, bartram2022untidy} and education practitioners expect an effective method for comprehending or learning the semantics of a program.
We note that the concerns of different roles are not strictly differentiated.
For example, data workers can also develop a wrangling program to find new insights on data.
In short, program visualization is usually used for debugging and education tasks~\cite{hoffswell2018augmenting}.}

Many debugging tools leverage visualizations to help developers identify and fix bugs. 
Some of them, such as Hdpv~\cite{sundararaman2008hdpv}, Heapviz~\cite{aftandilian2010heapviz}, and Anteater~\cite{faust2020anteater}, present task-specific or code-related information about the execution by giving a forest view. 
Others can reveal the runtime behavior, such as DDD~\cite{zeller1996ddd}, deet~\cite{hanson1997simple}, ZStep 95~\cite{lieberman1997zstep}, and VisuFlow~\cite{nguyen2018visuflow}. 
Whyline~\cite{ko2004designing} and Theseus~\cite{lieber2014addressing} introduce visualizations within integrated development environments, while FireCrystal~\cite{oney2009firecrystal} and Timelapse~\cite{burg2013interactive} focus on visualizing interactive behaviors on web pages.
Hoffswell \etal\cite{hoffswell2016visual} propose visual debugging techniques to inspect program states for reactive data visualization.
A number of works~\cite{harward2010situ, beck2013visual, beck2013situ, swift2013visual, hoffswell2018augmenting} leverage in-situ visualizations to display the program behavior.

\name{} can be used for debugging the process of data transformation.
However, our technique differs from prior work in two aspects.
First, instead of visualizing internal states or variables of programs, \name{} shows the semantics of code pieces, which involves input and output tables, the type of data transformation, and parameters of functions.
Second, data presented in the aforementioned approaches are generic types such as string and numbers.
On the contrary, data, in the context of data transformation, means 2-D data tables consisting of columns and rows.
The presentation of 2-D data tables is more challenging than generic data types.

Some program visualization systems are designed for education.
They intend to improve students' understanding of particular aspects of programs~\cite{urquiza2004survey}.
Online Python Tutor~\cite{guo2013online} is a web-based visualization tool that illustrates the runtime state of various data structures, which can be a valuable pedagogical aid for teaching Computer Science courses.
Algorithm visualization has been a hot research topic as having a significant impact on students learning behavior~\cite{grissom2003algorithm} and being promising for facilitating education~\cite{shaffer2011getting}. 
A variety of algorithm visualizations~\cite{hansen2002designing, demetrescu2002specifying, shaffer2007algorithm, Mike2014VA} depict program behavior on every step to facilitate understanding the program.
Some tools automatically convert source code to flow charts, including Visustin v7~\cite{oy2013visustin}, AutoFlowchart~\cite{AutoFlowchart}, code2flow~\cite{code2flow}, Flowgen~\cite{kosower2014flowgen}, and VizMe~\cite{cheon2019vizme}. 
The aforementioned approaches are explicitly designed for some algorithms or applications.
Nevertheless, none of them are proposed in the context of data transformation.
In this paper, we design and implement \name{} that the creation of evolution of data tables across a series of transformations.

\subsection{Data Wrangling}
Data wrangling is an arduous process of transforming, reformatting, and integrating data to make it more palatable for miscellaneous downstream purposes, including visualization and analysis~\cite{kandel2011wrangler}.
Many toolkits written in R (\eg, dplyr~\cite{wickham2021dplyr}, tidyr~\cite{wickham2020tidyr}) or Python (\eg, Pandas~\cite{reback2022pandas}) have been proposed to support the process. 
These toolkits provide excellent expressiveness for data workers to wangle data.
However, for data workers who are not proficient in R or Python, learning a new programming language or toolkits for wrangling tasks would spend substantial time and effort~\cite{shrestha2020here}.

To lower the barrier of data wrangling, various interactive systems and prototypes are proposed. 
Microsoft Excel, Tableau Prep Builder~\cite{Tableau2019}, and OpenRefine~\cite{OpenRefine} provide a menu-based GUI for users to iteratively clean, transform, and integrate data. 
Some systems embed a recommendation engine to suggest possible transformations. Data Wrangler~\cite{kandel2011wrangler,guo2011proactive} and its commercial successor Trifacta~\cite{Trifacta} recommend transformations based on users' manipulation.
The others, such as Foofah~\cite{jin2017foofah} and Wrex~\cite{drosos2020wrex}, borrow ideas from \textit{programming by example} that synthesizes code pieces for data transformation based on a small illustrative example provided by users.
Some systems support wrangling for graphs, websites, etc.
For example, Ploceus~\cite{liu2011network}, Orion~\cite{heer2014orion}, and Origraph~\cite{bigelow2019origraph} support graph editing and construction.
On the other hand, Vegemite~\cite{lin2009end}, Dataxformer~\cite{morcos2015dataxformer,abedjan2016dataxformer}, and WebRelate~\cite{inala2017webrelate}, are designed to transform data from different websites.

The aforementioned approaches assist data workers in conducting data transformations.
\name{}, on the other hand, targets presenting the process of data transformation.
Some tools, such as Tableau Prep Builder~\cite{Tableau2019}, OpenRefine~\cite{OpenRefine}, Data Wrangler~\cite{kandel2011wrangler,guo2011proactive}, and Trifacta~\cite{Trifacta}, record and present the process of data transformation using textual descriptions.
We argue that our visualization design is easier to understand and more effective than textual descriptions, and we report the comparison in Section~\ref{sec:userstudy} to justify the argument.

Kasica \etal\cite{kasica2020table} formed 21 types of operations based on a multi-table framework for data wrangling by two dimensions, \ie, three data types (rows, columns, and tables) and five operations (create, delete, transform, separate, and combine).
Furthermore, each type of operation is represented by an intuitive icon.
However, these icons are not mapped to data.
Inspired by these icons, we design our glyphs by supplementing them with parameters and additional types of visual channels to present the semantics of transformation operations.

\subsection{Provenance}

Provenance records the history of changes and advances during analysis~\cite{Ragan2016}.
Kandel \etal\cite{kandel2011research} emphasized the significance of capturing provenance from data quality operations and wrangling workflows when data workers share their data and scripts.

A number of works have been proposed to capture and visualize data provenance. 
For example, Tableau Prep Builder~\cite{Tableau2019} provides an icon for each operation in a data flow chart. Although these icons are easy to understand, they can not visualize the parameters of operations, such as the specificity of which tables/rows/columns are transformed and how.
\revise{By contrast, our glyph design can visualize both the type of data transformation and its parameters, which facilitates the comprehension of semantics because it reveals more details on data changes~\cite{liu2018steering}.} 
TACO~\cite{niederer2017taco} is a visual comparison tool for investigating the differences and changes between multiple tabular data over time. 
However, it focuses on quantitative homogeneous tables and does not support visualizing complex data transformations, including fold and unfold.
\name{}, on the contrary, focuses on visualizing the semantics of scripts and supports a wide range of data transformations mentioned by Kasica \etal\cite{kasica2020table}.
Some tools leverage animations to visualize data provenance.
Data Tweening~\cite{khan2017data} generates intermediate results for each data transformation in a SQL query session, facilitating the understanding and learning of complex transformations.
Datamations~\cite{pu2021datamations} explains the transformation steps of a data analysis pipeline by automatically generating a looping animated GIF from code.
Animation is useful for communication. 
However, the exploration of animation is slower as users often replay the animation dozens of times, and they can not control the animation at their own pace~\cite{fisher2010animation}. 
Additionally, those animation tools focus on presenting the data provenance of a single table.
In contrast, our work utilizes node-link graphs with glyphs to illustrate data provenance, which can better present the process of data transformations and portray the data provenance of multi-tables simultaneously.

%% file: src/requirement.tex
\section{Design Requirement}
\label{sec:req}
Our goal is to design a set of visual representations to help data workers understand and communicate a script of data transformation.
\revise{
To this end, we collaborate with two data analysts in a national research lab who have at least three years of expertise in data science.
Following Munzner's guidelines~\cite{munzner2009nested}, we conducted three rounds of interviews to iteratively extract design requirements.
Our interviews focused on their working scenarios, such as double-checking, where they are required to understand the semantics of wrangling scripts.
One major challenge is that they often need to recall or look up the usage and syntax of various functions.
One analyst reported, \textit{``I like Python. But sometimes I need to understand scripts written in R.''}
He added, \textit{``Cheat sheets are useful (to understand R functions) in many cases. I may also search in Google and RDocumentation\footnote{\url{https://www.rdocumentation.org/}} to understand advanced parameters.''}
However, there is a comprehension gap between the usage of functions and the semantics of practical code.
One analyst complained that he still needs to figure out how a line of code works on data after understanding the R function. 
In addition, the understanding of individual functions helps little in revealing the entire wrangling process.
In light of these complaints and feedbacks collected from interviews, we summarized the following design requirements.}
Particularly, R$1$ to R$4$ target the design of glyphs presenting individual data transformations, while R$5$ to R$7$ guide the design of \name{}.

\begin{compactenum}[\bfseries R1:]
\item 
\textbf{Present the Semantics:}\label{g:R1}
To help data workers understand a function, our visualization design should precisely present the semantics, including the function name, input, output, and parameters of a function.
As the number of functions could be large, designing visualization for each function may burden recognition.
Instead, we should present the type of data transformation to which the function belongs.
We distinguish between ``data transformation'' and ``function,'' as the former refers to a manipulation categorized in Kasica \etal\cite{kasica2020table} while function corresponds to a method in a programming language.

\item 
\textbf{Link with Data:}\label{g:R2}
When writing scripts, data workers usually need to ``look at" data tables by printing out a table or temporary results.
One analyst usually works with Jupyter\cite{jupyter}, and he commented, 
\textit{``I like to print out results to verify the operations.''}
Therefore, besides function-specific information, the visualization should reflect detailed information of a table, including content, shape, name, etc.

\item 
\textbf{Depict Necessary Information:}\label{g:R3}
Much information is involved in a function, such as function parameters, input and output data tables. 
We note that not all parameters are essential.
Similarly, when a table is large, illustrating all its content is impossible and unnecessary.
As a result, we should elicit and encode critical information from a function and representative content in a table.

\item 
\textbf{Keep Encoding Consistent:}\label{g:R4}
Glyphs in \name{} should have consistent visual encodings.
When visualizing a sequence of functions, consistent visual encoding facilitates understanding each data transformation and the entire procedure. 

\item 
\textbf{Reveal Table Provenance:}\label{g:R5}
Data tables are evolved and correlated through functions. 
For example, an output table of a data transformation may serve as input for another, and so forth.
Data provenance records how data was created and changed, which is significant in tracing data processing changes back to their original sources~\cite{bors2019capturing}.
As one analyst noted, \textit{``Some operations (such as join) rely on multiple tables. 
Displaying how tables are correlated is useful in debugging.''}
Hence, the visualization should provide an overview of the entire data provenance.

\item 
\textbf{Dig into Details:}\label{g:R6}
Data workers usually need to validate a series of functions.
Hence, they seek to grasp detailed information on each data transformation.
Our visualization should allow users to switch between an overview and a detailed view of individual transformations.

\item 
\textbf{Independent of Programming Languages:}\label{g:R7}
Data workers may use various toolkits for data transformation, such as dplyr~\cite{wickham2021dplyr} in R and Pandas~\cite{reback2022pandas} in Python. 
To ensure generalizability, our visualization design should be independent of toolkits and programming languages.

\end{compactenum}

%% file: src/glyph.tex
\section{Design of Glyphs}
\label{sec:design_space}

Guided by the aforementioned design requirements, we design a collection of glyphs that presents the semantics of functions.
To answer questions like, ``which information shall we encode'' and ``which visual channel can be mapped,'' we structure the design space by two primary dimensions, \ie, the type of parameters and potential visual channels.

\subsection{Parameters Space}
\label{subsec:space_params}
The analysis of toolkits helps us identify key information that should be encoded in the design of glyphs \kai{(R\ref{g:R1}, R\ref{g:R3})}.
In this paper, we focus on three packages in R and Python, \ie, dplyr, tidyr, and Pandas, because they all target data transformation and are open-source in nature.
We have examined $160$ functions in total, where $84$ are from dplyr, $23$ from tidyr, and $53$ from Pandas.
We read the official documentation of these functions, reveal semantics under different parameters, and map them to the type of transformations.
Moreover, we run these functions with different parameters to understand how parameters affect the transformation results.
Finally, we categorize six key parameters that should be mapped to visual channels.

\textbf{Function Name} reflects which operation does the function targets. 
Since the name does not have a one-to-one mapping with data transformations, the information, in some cases, benefits little in communicating the semantics of a function. 
For example, the \textit{select} function in dplyr can be mapped to three different transformations, \ie, \textit{Delete Columns}, \textit{Rearrange}, and \textit{Transform Columns}, depending on parameters and data tables.
We do not emphasize function names in our visualization design. 

\textbf{Data Tables} are described as variables in the script and are input and output of a function.
We identify a variable string as a table name.
Data tables are stored in a well-designed data structure, \eg, data.frame in R or DataFrame in Python, and can vary in dimensions 
In this paper, we focus on 2-dimension tables, which are collections of rows and columns.

\textbf{Explicit Columns/Rows} are the columns/rows explicitly mentioned as parameters in functions.
Explicit columns are usually referred to using column names, while explicit rows are mentioned using row indexes.  
Taking the statement as an example, \textit{tree2=arrange(trees, Girth)}, the parameter \textit{Girth} is the name of an explicit column in the table \textit{trees}. 
Another example is that, in the statement \textit{mtcars\_temp=slice(mtcars, 1, 5)}, the parameters \textit{1, 5} are two-row indexes of the table \textit{mtcars}.
The quantities of explicit columns and rows are usually limited.
Since they are key information in a function, they should be illustrated and highlighted in the glyph design.

\textbf{Implicit Columns/Rows} are not listed as parameters in a function. 
Rather, they are selected in the data transformation based on filtering criteria. 
For example, when deleting duplicate rows, rows with identical values are compared and filtered.
The presentation of implicit columns/rows is beneficial for understanding data transformation. 
Because the volume of implicit elements is usually large, depicting all these is virtually impossible. 
As a result, we should select and encode representative ones in the glyph.

\textbf{Contextual Columns/Rows} are not involved in a data transformation. 
Specifically, they do not meet filtering criteria are not selected during transformation.
Similar to explicit and implicit elements, we argue that context is also useful for communicating data transformation. 
For example, contextual columns keep unchanged when deleting a column to show a contrast to the deleted one~\cite{kasica2020table}.
Similar to implicit elements, we should encode a limited number of contexts.

\textbf{Transformation Parameters} 
Beside the type of data object, a function usually includes a number of parameters to precisely acknowledge function details.
They can be inline functions (\eg, sum, min, regular expression, etc.), mathematical operators (\eg, ``+,'' ``-,'' etc.), or transformation-specific identifiers (\eg, separator in \textit{separate} and \textit{unite}).
The visualization of these parameters is critical to revealing subtle differences among transformations.

\subsection{\kai{Design Rationale}}
\label{subsec:space_channels}

Kasica \etal\cite{kasica2020table} structured a multi-table framework for data wrangling.
The framework includes $15$ categories of transformations, and each may contain several subtypes.
For example, based on whether the operation modifies table schema, \textit{Transform Tables} includes two subtypes, \ie, \textit{Rearrange} and \textit{Reshape}.
Further, they designed icons for $21$ (sub)types of data transformations. 
These icons are intuitive and inspiring and provide a good starting point for our glyph design.
We distinguish between an icon and a glyph, in which the former is a visual representation only and irrelevant to data; in contrast, the latter, which is widely used in various tasks to represent multidimensional data~\cite{ying2021glyphcreator, wang2021tac, tang2021videomoderator, deng2021compass}, maps data to visual channels such as color, size, etc.
By analyzing all icons and the parameter space, we distill the following design guidelines in creating our glyph collection.

\begin{figure}[!t]
    \centering
    \includegraphics[width=\linewidth]{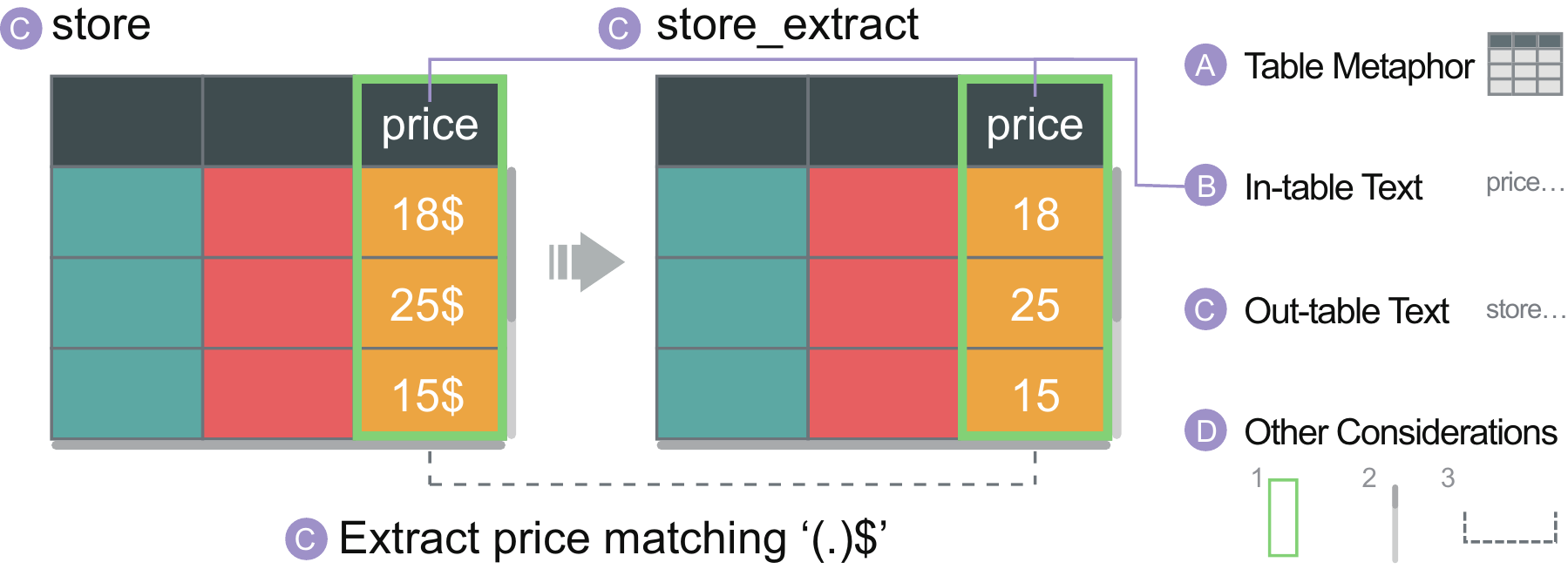}
    \caption{We use \textit{Transform Columns} as an example to showcase different visual channels.}
    \label{fig:visual_channels}
    \vspace{0mm}
  \end{figure}

  \begin{figure*}[!t]
    \centering
    \includegraphics[width=\linewidth]{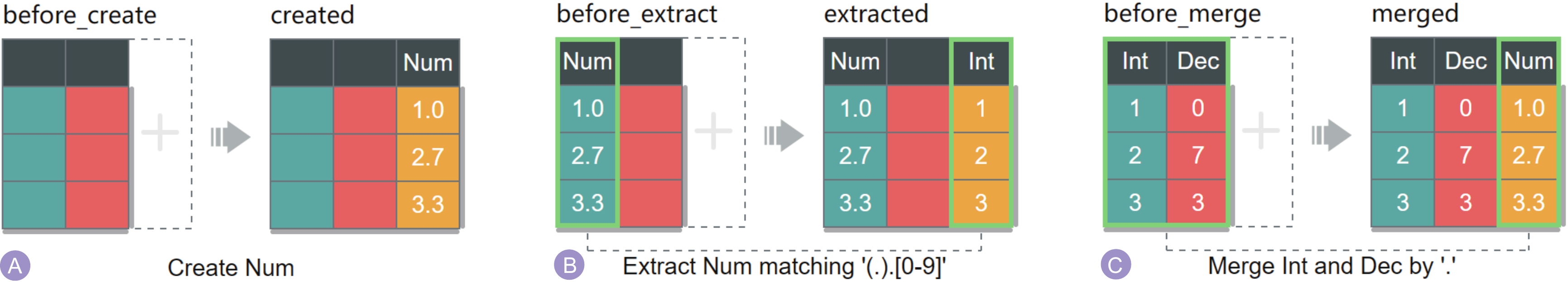}
    \caption{ By depicting in-table and out-table text, a glyph can differentiate different types of transformations. Taking \textit{Create Columns} as an example, (a) shows the creation by filling in values manually, (b) shows extracting values from existing columns, and (c) depicts merging values from existing columns.
    }
    \label{fig:in_table_text}
  \end{figure*}

\textbf{Input and Output Tables:}
Each icon designed by Kasica \etal is composed of three main parts, \ie, an input table, an output table, and an arrow indicating the transformation.
We follow this metaphor (Figure~\ref{fig:visual_channels}(a)) in designing our glyph collection because these are necessary for presenting a transformation \kai{(R\ref{g:R1})}.
In the following description, we use ``data'' to indicate data tables and use ``table'' to refer to the table metaphor in a glyph.

\textbf{Table Shape:}
The shape of a table is affected by the number of explicit, implicit, and contextual columns/rows \kai{(R\ref{g:R1}, R\ref{g:R3})}. 
All explicit entities are depicted in the table due to their importance.
For implicit entities, we selectively choose one that is representative. 
In some cases, ``one'' means ``one pair''. For example, when depicting ``remove duplicate rows'', we select two identical rows as implicit entities. 
Contextual entities are displayed for two reasons. 
First, it helps to present the semantics of a transformation by posing a contrast to explicit/implicit entities.
Second, it retains the table metaphor. For example, in ``Create Table'', all entities are contextual. 
We demonstrate $3\times 3$ contextual cells to indicate an empty table.
For other transformations, context is limited to one or two columns/rows.

\textbf{Cell Color:} 
Color encoding is \revise{meaningful} in Kasica \etal\cite{kasica2020table}.
It is designed for distinguishing cell types (\eg, title cells are white while content cells are colored), indicating the type of data object (\eg, column and row), depicting unchanged columns/rows, and presenting correlated columns/rows (\eg, the icon for \textit{Interpolate}).
The color encoding of our glyph is primarily borrowed from Kasica \etal. 
Further, we extend prior work from two aspects. 
First, we use white color to represent empty cells. At the same time, title cells are colored dark gray. 
Second, we use striped cells to depict those with an empty or blank string \kai{(R\ref{g:R4})}.

\textbf{Out-table Text:}
Some text is displayed outside the table. 
For example, for transformations targeting specific rows by row index, we present row index aside from the table \kai{(R\ref{g:R3})}.
Besides, we present table names and the type of transformation that a function belongs to. 
Following the text in Trifacta~\cite{Trifacta}, we present textual information below the input and output tables to describe the semantics of transformations (Figure~\ref{fig:visual_channels}(c)) \kai{(R\ref{g:R1})}.

\textbf{In-table Text:}
Presenting data content is critical to assisting data workers to understand a function \kai{(R\ref{g:R1}, R\ref{g:R2})}. 
Due to limited glyph size, only contents in explicit and implicit columns/rows are depicted in the glyph (Figure~\ref{fig:visual_channels}(b)).
Usually, it is not possible to present all values in these elements. 
Hence, we randomly sample values from data.
By encoding in-table text, as well as out-table text, we are able to distinguish data transformations with a subtle difference.
For example, there are four common subtypes of transformations for \textit{Create Columns}\cite{kasica2020table}, \eg, creating manually (Figure~\ref{fig:in_table_text}(a)), mutating from other columns (example of \textit{Create Columns} in Figure~\ref{fig:operations_glyphs}), extracting substring of one column (Figure~\ref{fig:in_table_text}(b)), and merging multiple columns (Figure~\ref{fig:in_table_text}(c)).
Another type of information shown in a table is special symbols.
For example, we use \includegraphics[height=\myMheight]{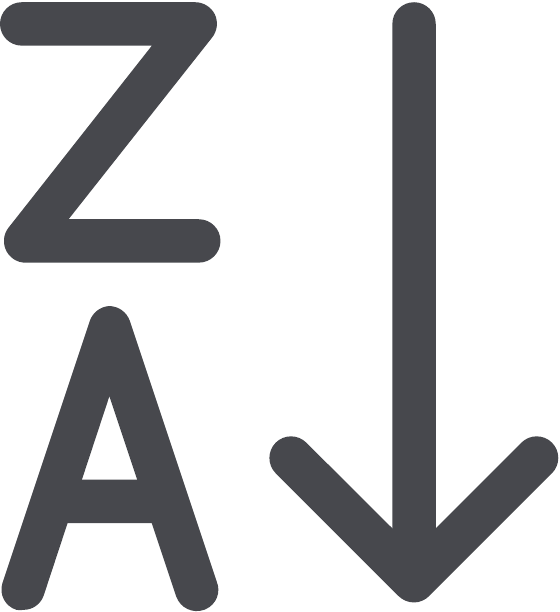} and \includegraphics[height=\myMheight]{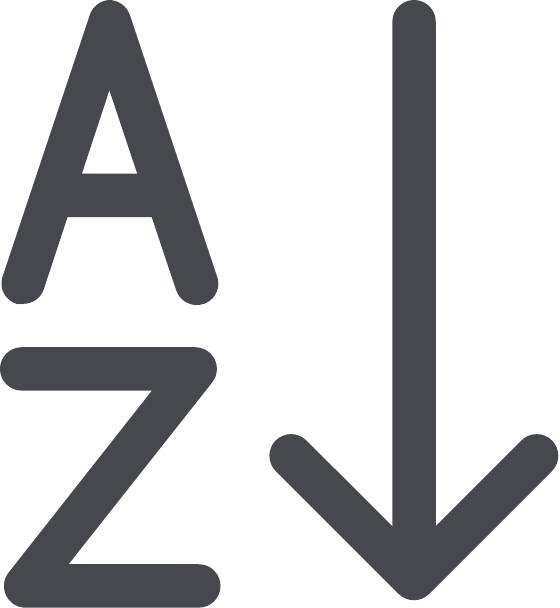} to illustrate sorting in descending and ascending order, respectively.

\textbf{Other Considerations:}
Besides the aforementioned guidelines, we explore visualization techniques that enhance the perception of data tables and transformation (Figure~\ref{fig:visual_channels}(d)).
First, to indicate the shape of data, we design both horizontal and vertical scroll bars in glyphs.
The size of scroll bars is proportional to the shape of data tables.
We acknowledge that this design cannot show a precise number of columns/rows.
Instead, it informs that the glyph presents a portion of an entire data table \kai{(R\ref{g:R2})}. 
Second, to emphasize the change of a table \kai{(R\ref{g:R1})}, we highlight the correlation between explicit columns in input and output tables. 


\subsection{Results}
\label{subsec:glyph_results}
Following the aforementioned design space, we derive $21$ glyphs for data transformation. 
Besides, we create glyphs for two more transformations.
First, in Kasica \etal \cite{kasica2020table}, \textit{Fold} and \textit{Unfold} share one icon with different arrow directions.
We distinct the two operations with two glyphs.
Second, we add a glyph for \textit{Rearrange Columns} because it is triggered by a popular function, \textit{select}, in dplyr.
To save space, Figure~\ref{fig:operations_glyphs} illustrates $15$ out of $23$ glyphs, and the full glyph collection can be found in the supplemental material.

%% file: src/pipeline_design.tex
\section{Design of \name{}}
\label{sec:pipeline_design}
\begin{figure}[!ht]
  \centering
  \includegraphics[width=\linewidth]{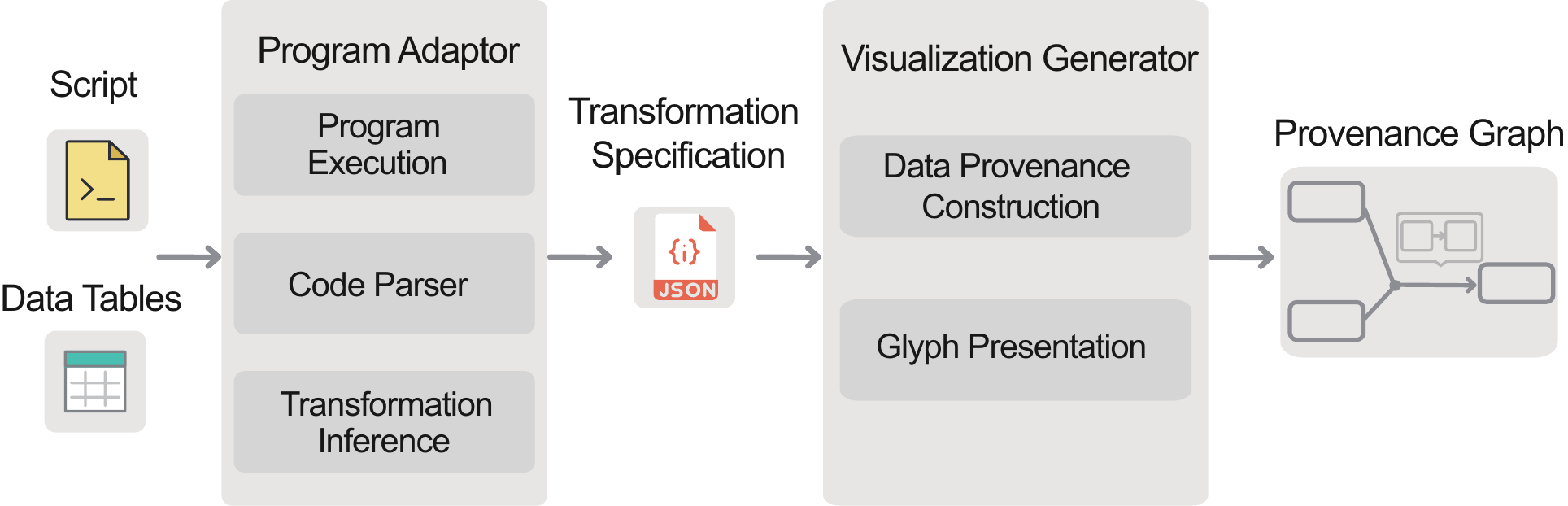}
  \caption{The architecture of \name{} consists of two major modules: Program Adaptor and Visualization Generator.
  The Program Adaptor accepts a script and data tables as input and outputs a collection of transformation specifications.
  The Visualization Generator generates table provenance by utilizing the specifications. 
  }
  \label{fig:pipeline_architecture}
  \vspace{-0.23cm}  
\end{figure}

In this section, we present the design of \name{}, a pipeline that accepts data tables and a piece of code as input, and results in a visual representation to show the entire procedure of data transformations.
Code pieces may contain complex control flow, including a conditional statement, loops, function definition, etc.
In this paper, we limit the scope to code consisting of assignment statements only.
Though some modules are implemented based on specific programming languages and toolkits, we argue that the design of \name{} can be applied to different programming languages.
In the presentation, we use dplyr, a toolkit of R, as examples by default.
Figure~\ref{fig:pipeline_architecture} shows the architecture of \name{}, which consists of two core modules, \ie, Program Adaptor and Visualization Generator.

\begin{figure*}[!h]
  \centering
  \includegraphics[width=\textwidth]{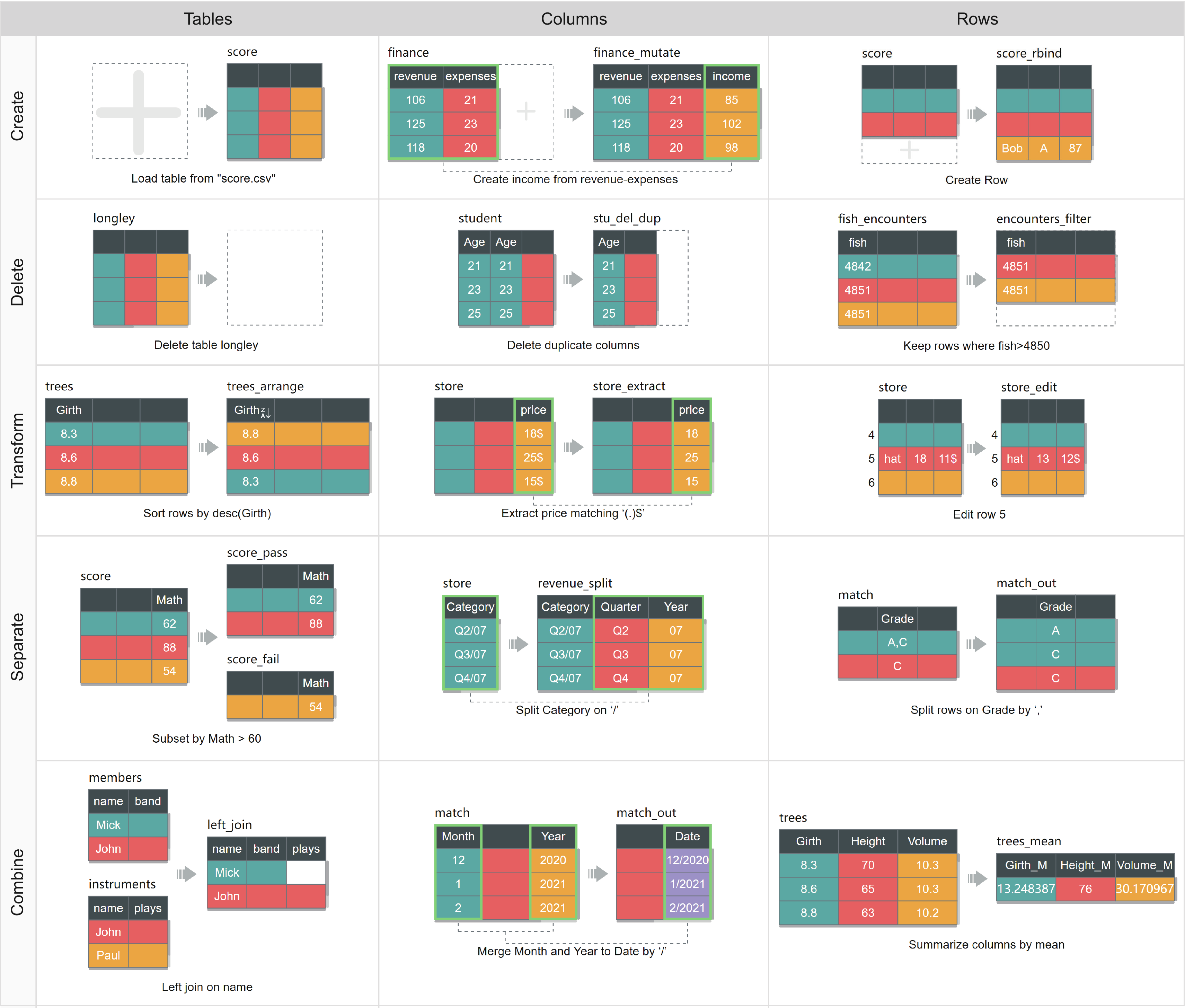} 
  \caption{
 Following Kasica \etal\cite{kasica2020table}, we display $15$ out of $23$ transformations by two dimensions, \ie, the type of data object and five operation categories.
 All glyphs are generated based on real data tables and functions from tidyr~\cite{tidyr_package} and dplyr~\cite{dplyr_package}.
 The entire glyph collection can be found in the supplemental material, \kai{which is available online at \url{https://github.com/xkKevin/Somnus}}. 
  }
  \label{fig:operations_glyphs}
  \vspace{-0.23cm}  
\end{figure*}

\subsection{Program Adaptor}
Program Adaptor aims to generate a series of transformation specifications given data tables and a script.
Though we implement an adaptor for each programming language, all adaptors share three common steps, \ie, Program Execution, Code Parser, and Transformation Inference.
The descriptions of this module are independent of programming languages \kai{(R\ref{g:R7})}.

\subsubsection{Program Execution}
After a script and data tables are fed into the Program Adaptor module, the script will be executed using an interpreter based on programming languages.
The primary goal of this step is to obtain input and output data for each function, which is beneficial for 1) providing data value when plotting glyphs \kai{(R\ref{g:R2})} and 2) inferring the type of data transformation for each function \kai{(R\ref{g:R1})} (see Transformation Inference for details). 
The Program Execution steps automatically insert statements for importing necessary libraries to interpret and execute the script correctly.

\subsubsection{Code Parser}
Code Parser accepts a script as input and parses each line of code to obtain 1) the name of the input and output table, 2) function names, and 3) parameters of functions \kai{(R\ref{g:R3})}.
In some toolkits, a transformation can be invoked through various approaches.
For example, given a data table, named ``tbl,'' containing three columns in order, \eg, ``column1'', ``column2'', and ``column3''. 
Assume ``tbl'' is stored as DataFrame in Python, \textit{Rearrange Columns} can be expressed as \textit{pandas.DataFrame(tbl, columns=[`column2', `column1', `column3'])}.
Also, the same transformation can be achieved by \textit{tbl[[`column2', `column1', `column3']]}.
In the current implementation of \name{}, we only support statements that have explicit function names and input and outputs.

Similar to Program Execution, results generated by Code Parser are critical to inferring the type of transformations \kai{(R\ref{g:R1})}.
Besides, tracing the input and output tables helps to construct the provenance of data.

\subsubsection{Transformation Inference}
\label{subsec:transformation_inference}
After parsing individual functions and their input and output tables, we build a mapping between functions and the type of transformations \kai{(R\ref{g:R1})}.
In most cases, one function is mapped to one data transformation. 
We create rules for mapping the name and parameters of a function to one type of transformation. 
For example, we map \textit{filter} to \textit{Delete Rows}, \textit{separate} to \textit{Separate Columns}, and \textit{count} to \textit{Summarize}.
However, function information is not enough in some cases.
For example, given the data table (``tbl'') mentioned above, the statement \textit{select(tbl, ``column3'', ``column1'', ``column2'')} equals to \textit{Rearrange Columns}.
On the other hand, if the input table has four columns in order, \eg, ``column3'', ``column1'', ``column2'', and ``column4'', the same statement results \textit{Delete Columns} as ``column4'' is omitted in the output table.
In this case, we derive the type of transformations by comparing the input table with the output.

We note that some functions involve a sequence of transformations.
Taking the data table (``tbl'') as an example, the function \textit{select(tbl, column1, column4 = column2)} first performs \textit{Delete Columns} by deleting ``column3''. 
Then, it \textit{Transform Columns} by renaming ``column2'' to ``column4''.
In these cases, we identify multiple transformations for a function.
One challenge is to obtain input and output tables for each transformation, which are critical to glyph generation.
The current prototype establishes rules and replaces a function with multiple ones, where each corresponds to a transformation. 
Then, the entire script is executed again to derive the input and output data tables.

To save screen space, some functions can be grouped and merged.
For example, the two functions, \eg, \textit{rename(tbl, column4 = column1)} and \textit{rename(tbl, column5 = column2)}, can be combined into one \textit{rename(tbl, column4 = column1, column5 = column2)}.
In such cases, we depict the two functions using one data transformation.
The current prototype supports the combination of consecutive functions in three cases, \ie, \textit{Rename Columns}, \textit{Delete Columns}, and \textit{Delete Rows}.
We plan to investigate more rules for combining the semantics of functions in future research.

\subsubsection{\revise{Failure Modes}}
\revise{To improve reliability, \name{} is able to deal with five types of failure modes.
First, if the script contains operations that are unsupported, \name{} can compile and run the script properly. 
However, there are no glyphs for these operations.
Second, if the function is not supported in the Program Adaptor, such as \textit{drop\_na} in tidyr, \name{} shows the function name only. 
Third, for operations that are not function-based, \eg, ``\textit{df = df[df.col1 \textgreater 0]}'' in Pandas, \name{} displays nothing on the edges.
Fourth, \name{} decomposes operations involving many columns and rows into multiple ones and visualizes them with a sequence of glyphs, as described in Section~\ref{subsec:transformation_inference}.
Finally, \name{} does not fully support the parsing of non-assignment statements, such as loops or conditional statements.
We acknowledge that this policy may hinder data workers from understanding the logic of the entire script.
For example, given a conditional statement like \textit{``if CONDITION then OPERATION1 else OPERATION2''},
\name{} displays only one glyph representing either \textit{OPERATION1} or \textit{OPERATION2} based on the results of Program Execution.}

\subsection{Visualization Generator}
The result of Program Adaptor is a set of transformation specifications accompanied by input and output data tables.
This module aims to generate visual representations depicting both an overview and detailed information for transformations.

\subsubsection{Constructing Data Provenance}
\label{subsubsec:provenance_graph}
\begin{figure}[ht]
  \centering
  \includegraphics[width=\linewidth]{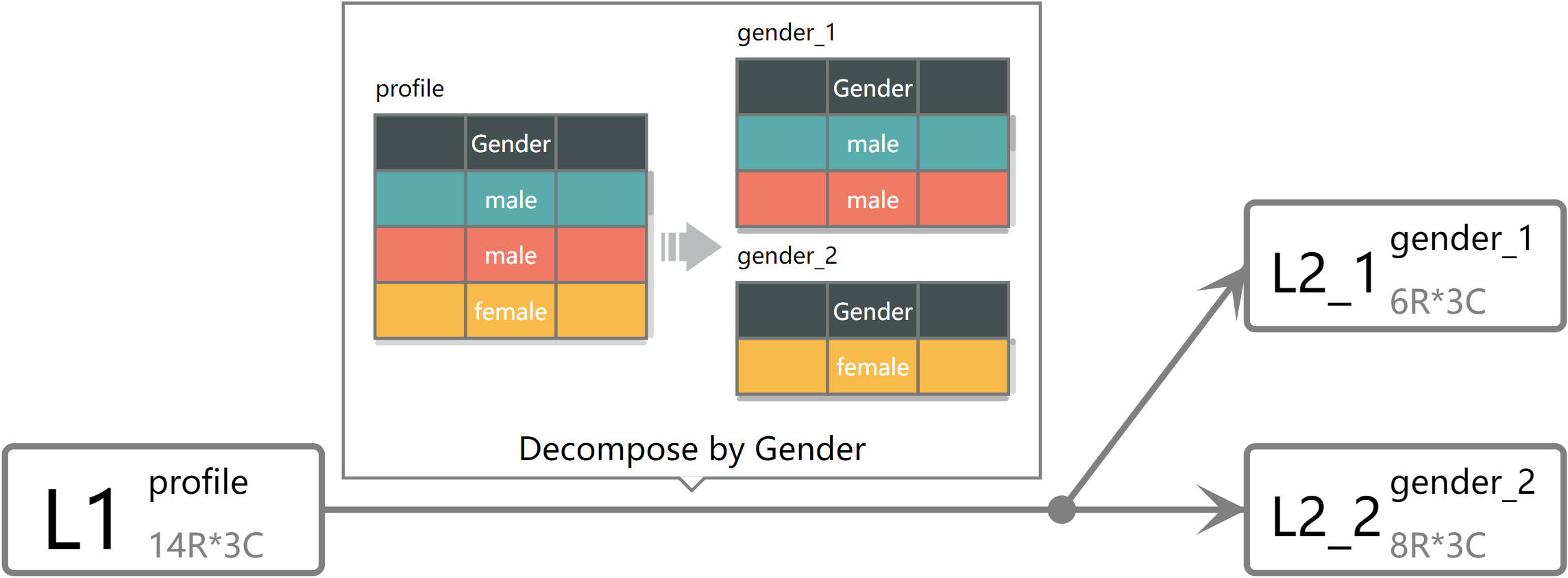}  
  \caption{
    The transformation, \textit{Decompose}, results in a divergence edge, which is generated by two statements, \textit{profile = read.csv("profile.csv")} and \textit{gender = group\_split(profile, Gender)}.
  }
  \label{fig:transformation_edges}
\end{figure}

Data provenance can be formed as a graph, where nodes are data tables and edges are data transformations \kai{(R\ref{g:R5})}. 
The provenance graph is layered, and we leverage the Eclipse Layout Kernel~\cite{ELK2021} for positing graph nodes.
As shown in Figure~\ref{fig:transformation_edges}, each node is rectangular, and we depict useful table information in each node, including the line index where the table is created, the table name, and the size of the table.

Edges connecting nodes are data transformations.
According to the number of input and output tables, we categorize the edges into three types.
In most cases (Figure~\ref{fig:reverse_engineering}(d)), a transformation accepts a table as input and outputs a transformed one.
We depict the edge as a directed line.
Second, some transformations merge multiple data tables into one, such as \textit{Extend}, \textit{Supplement}, and \textit{Match}. 
These transformations are shown as convergence edges (Figure~\ref{fig:teaser}(c)).
Similarly, transformations that result in multiple output tables from one input are depicted as divergence edges, as shown in Figure~\ref{fig:transformation_edges}.

\subsubsection{Presenting Glyphs}
To present the details of data transformations, we depict glyphs aside from each edge in the provenance graph \kai{(R\ref{g:R6})}.
Since glyphs may contain in-table and out-table text, they are placed horizontally without rotation for better readability.
For functions that are not supported by our glyph collection, no glyph is displayed.

\subsection{Implementation}

\name{} is implemented as a web-based client-server system.
The backend is implemented using flask, while the frontend is built in Vue.js and D3.js~\cite{bostock2011d3}.
The web interface consists of four panels, \ie, a Data Panel, a Script Panel, a Table Panel, and a Graph Panel.
The Data Panel (Figure~\ref{fig:reverse_engineering}(a)) allows users to upload their input tables as needed.
Users need to select a programming language and copy-paste a piece of data wrangling code to the Script Panel (Figure~\ref{fig:reverse_engineering}(b)).
Then the backend runs based on the input tables and the script provided by the user.
The Graph Panel (Figure~\ref{fig:reverse_engineering}(d)) displays table provenance (R\ref{g:R5}), while the Table Panel (Figure~\ref{fig:reverse_engineering}(c)) is used to show the intermediate tables generated in the process of data wrangling (R\ref{g:R2}).
To assist the investigation of lengthy table provenance, the Graph Panel supports zooming and panning \kai{(R\ref{g:R6})}. 

Interactions across panels are integrated to facilitate the exploration among script, data tables, and data provenance.
First, when a user clicks on a node (\ie, data table) in the provenance graph, the Table Panel displays the detailed table.
Similarly, when clicking on edge (\ie, transformation) in the graph, its function in the Script Panel is located and highlighted, and vice versa.

We focus on data wrangling from two popular programming languages, \ie, R and Python.
Specifically, the current prototype supports $25$ commonly used functions from tidyr\cite{tidyr_package} (\eg, \textit{separate}, \textit{gather}, \textit{spread}, etc.) and dplyr\cite{dplyr_package} (\eg, \textit{filter}, \textit{select}, \textit{mutate}, etc.), and ten functions from Pandas~\cite{reback2022pandas} (\eg, \textit{pandas.unique}, \textit{pandas.merge}, \textit{pandas.concat}, etc).

%% file: src/evaluation.tex
\section{User Study}
\label{sec:userstudy}
To assess the effectiveness of the visualization design, we conducted a controlled study centered on two high-level questions: 
1) does the glyph design improve user efficiency in comprehending the semantics of data wrangling?
and 2) does the provenance graph facilitate the understanding of data dependencies?
We ran the evaluation using real-world data tables and scripts written in the R programming language.
All documentation, including scripts, data tables, questions, etc., are provided in the supplemental material.

\subsection{Participants and Apparatus}
We recruited $20$ volunteers ($4$ females and $16$ males) aged $22$ to $35$ ($\mu =25.45$, $\sigma =3.33$).
The majority of participants ($15/20$) were postgraduate students majoring in Statistics or Computer Science, while the others worked as data analysts or algorithm engineers in a national research lab.
They were all proficient in programming using Python, JAVA, or Javascript and had experience in data transformation.
In addition, to ensure that all participants had difficulties understanding scripts, we only recruited those who had not written a line of code in R.
Participants completed the study using a desktop computer (3.20GHz 8-Core Intel Core i7, 32 GB memory) with a 27-inch monitor ($3840\times 2160$ resolution) and an external mouse and keyboard, and the study was distributed through Google Chrome on a Windows 10 machine.

\subsection{Techniques}
To our knowledge, no prior work targets visualizing the semantics of data transformation.
Hence, we compared the visualization design with textual description derived from a commercial data wrangling system, Trifacta~\cite{Trifacta}.
Some description was not directly supported by Trifacta, such as mapping the values from one column into another (a subtype of \textit{Transform Columns}). 
In these cases, we generated text by combining the descriptions of two transformations, \eg, create \textit{a new column} from \textit{original columns} and delete \textit{the original ones} (italic text will be replaced by column names).
To align with \name{} in describing a sequence of transformations, we included additional information in the textual description, including line index, the shape of data tables, and the output table name of a function. 

The design of our glyph collection contained textual information describing the type of transformation in the glyph.
We noted that the comparison between pure text and visualization with text would be unfair.
To evaluate the effectiveness of visual representation, we removed the textual description of glyphs in the study.
We envisioned that our glyph design and \name{} would be more effective by including text.

\subsection{Tasks and Design}
We performed a within-subject design with two experimental techniques and ten experimental tasks.
To address any memory learning effects, we created two different sets of ten tasks.
The orders of the two techniques and the task sets were counterbalanced using a Latin square. 
Within each technique, participants completed ten tasks which were shown in a fixed order.
\kai{Thus, the whole study contained $2$ techniques $\times$ 2 sets $\times$ $10$ tasks $=$ $40$ trials.}
Each task trial included a piece of code, a visual or textual explanation for the code, and a multiple-choice question where each question had one or more correct answers.
In addition, the study system provided data tables and documentations of functions for reference.
\revise{We chose the multiple-choice test for evaluation because it could increase participants' confidence in completing tasks and be more convenient for statistical analysis of test results over the constructed-response test~\cite{kuechler2003well, simkin2005multiple}.
In our study, all questions and choices are carefully designed from varying perspectives of table changes and dependencies to answer the above two high-level questions.
However, we do not guarantee they can genuinely measure participants' understanding as comprehension is abstract and hard to access directly.}


The ten study tasks consisted of five function understanding tasks (\textit{Func}) followed by five script understanding tasks (\textit{Script}).
Moreover, each question was required to select all correct choices.
\textit{Func} focused on the semantics of individual functions, including the output tables and operations.
Example choices were statements such as, \textit{``The output table has a different number of rows with the input table''} and \textit{``This function renames the column A to B''}.
\textit{Script} focused on the understanding of data provenance through a sequence of functions.
Example questions were, \textit{``Which data tables contribute to the creation of table A?''} and \textit{``How many data transformations are performed from table A to B?''}
We carefully designed the task questions and choices to avoid ambiguous answers and maintain the same difficulty across task sets.
To assist the exploration of data provenance, the study system supports zooming and panning.
A participant answered a question correctly if and only if all correct options were selected.

\subsection{Data}
To evaluate how our visual design performed in real-world scenarios, we selected candidate code pieces and data tables provided by Kasica \etal\cite{kasica2020table}, collected from Github~\cite{beecycles2018, Baltimore2020}.
We focused on functions belonging to two toolkits, \ie, dplyr~\cite{dplyr_package} and tidyr~\cite{tidyr_package}. 
We randomly chose a set of statements for \textit{Func} and consecutive code pieces for \textit{Script}.
Due to the limitation of the programming adaptor, some statements, or code pieces, were inadequate for our studies, such as those without explicit function names and input tables.
Hence, we replaced these statements with their functional alternatives.
In addition, we removed comments to avoid misleading.

We created two datasets, and each corresponded to one task set.
To maintain the same difficulty level across the two datasets, we established three rules in dataset creation.
First, both datasets contained the same number of \textit{Combine Tables} and \textit{Separate Tables}. 
Because these transformations involved more data tables compared to the rest, they might pose challenges in understanding table provenance.
Second, each dataset included the same number of functions for \textit{Func} and \textit{Script}.
In our study, five functions were applied to \textit{Func}, and nine functions were used for \textit{Script}.
To keep the datasets distinct, functions in one dataset could not be reused in the other.
However, exceptions existed for two functions, \ie, \textit{read.csv()} and \textit{group\_by()}, to ensure that code pieces can be correctly interpreted.
Specifically, \textit{read.csv()} loaded data at the beginning of code pieces while \textit{group\_by()} served as a prerequisite for other functions, such as \textit{summarise()} and \textit{mutate()}, to achieve some transformations.

\begin{figure*}[tb]
    \centering
    \includegraphics[width=\linewidth]{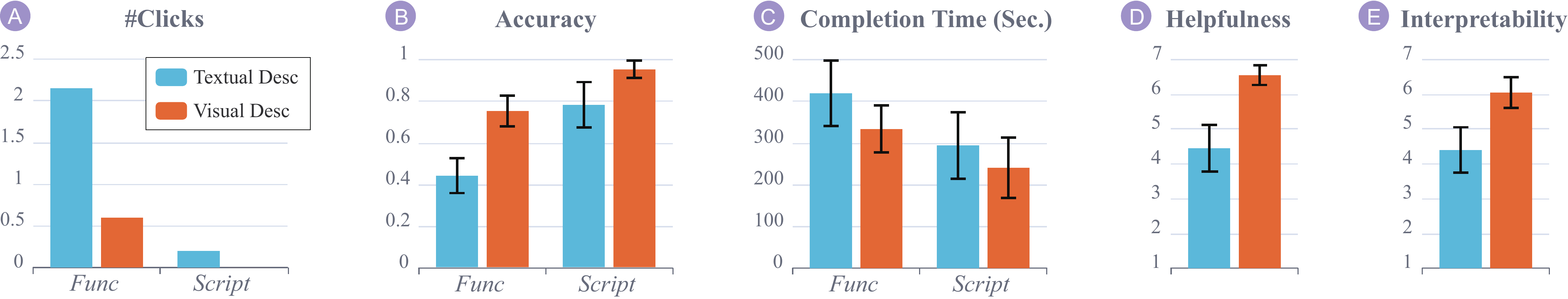}
    \caption{
        \kai{Among participants,} (a) shows the average number of clicks in terms of two task sets and two techniques, 
        (b) and (c) are the average accuracy and completion time. (d) and (e) display user ratings in terms of helpfulness and interpretability. The error bars indicate the 95\% confidence interval.} 
    \label{fig:qualitative_result1}
    \vspace{-0.15cm}  
\end{figure*}


\subsection{Procedure}
The study began with a brief introduction to data transformation.
Then, we collected demographic information of each participant, including the experience of programming, age, occupation, etc.
Prior to the main experiment for each technique, participants performed ten training tasks with a separate dataset.
During training sessions, participants were instructed to think aloud, and the experimenter helped answer questions and overcome difficulties.
We reminded participants that they could always skip a task when they were not confident about the answer.
In the main experiment, participants were asked to complete ten tasks with each technique.
The data tables and documentations of functions provided by the study system were folded by default.
Participants could click to expand this information for reference. 
Our system recorded the clicks for further analysis.
In addition, the system recorded task completion times and participants’ answers.
After the main experiment for each technique, participants were asked to rate the usefulness and intuitiveness of the technique using a seven-point Likert scale.
After the study, a semi-structured interview was conducted to collect their feedbacks.
We took notes during the whole session.
Each participant took approximately one hour to finish the study and received $10$ dollars as compensation.

\subsection{Quantitative Results}
\textbf{Accuracy: }
\revise{The individual answer-level results of each question across the two techniques are provided in the supplemental material.}
We \revise{found} that the accuracies of Q1 in \textit{Func} Set1 and Q4 in \textit{Func} Set2 were very low in both two techniques.
There were two possible reasons for the results.
First, the semantics of some \textit{Combine} operations like \textit{Summarize} and \textit{Supplement} were hard to describe, which involved the rule of combination, and the number of rows and columns.
Second, participants seldom chose the NOTA option \revise{(\ie, \textit{``None of the above''})}, which was the correct answer for the two questions.
As P16 explained, \textit{``When I am not sure about the answer, I tend to choose the one that seems correct (instead of NOTA).''}
Besides, visual descriptions' accuracy was significantly higher than that of textual descriptions (see Q5 in \textit{Func} Set2 and Q4 in \textit{Func} Set1), which indicates that our visual design is superior to text in describing complex operations, including \textit{fold} and \textit{left join}.
Figure~\ref{fig:qualitative_result1}(b) depicts the results with a 95\% confidence interval.
On average, participants achieved a much higher accuracy with visual description (\kai{$\mu =0.85$, $\sigma =0.16$}) than with textual description (\kai{$\mu =0.61$, $\sigma =0.27$}) for all tasks.
Especially for \textit{Func}, participants got an average accuracy of \kai{$44\%$ ($\sigma =0.18$)} with the baseline technique. 
And they achieved an accuracy of \kai{$75\%$ ($\sigma =0.16$)} using our approach.
In our study, textual description described what the transformation was.
However, it helped little in communicating how the transformation performed.
As P9 commented, \textit{``Though the text told me that the function performs left join, I do not know exactly how left join works.''}
On the contrary, visual description helped participants understand the semantics of transformations that they were unfamiliar with.

\textbf{Completion time:}
We performed an independent-samples t-test with a null hypothesis that the participants took the same amount of time finishing tasks with each technique.
We found a marginally positive effect of our approach with which participants completed \textit{Func} faster than the baseline technique ($p < 0.1$).
We also ran a paired two-sample Wilcoxon signed-rank test to identify whether the presentation order of two techniques affected the task completion time.
The results indicated \kai{no} significant effect of the order on the completion time for \textit{Func} ($p = 0.1536$) \kai{while} a notable significant effect for \textit{Script} ($p = 0.0083$).
That is, participants, performed faster using the later technique \kai{in \textit{Script}}.

\textbf{Number of clicks:}
In \textit{Func}, participants expanded function documentation and data tables $2.15$ times using textual descriptions. 
On the contrary, they clicked $0.6$ times using our approach.
Compared to the baseline approach, participants sought less information in completing the tasks. 
This result indicated that our approach helped participants understand the semantics of transformations with necessary visual encoding. 
The number of documentation and table clicks was much fewer in \textit{Script}, which might be because this information helped little in script understanding tasks.

\textbf{Preference:}
For comparing the helpfulness and interpretability of the two techniques across ten tasks, we ran Mann-Whitney's U tests to evaluate the difference in the responses of our seven-point Likert Scales. 
We found our technique ($\mu =6.55$, $\sigma =0.61$) was significantly more helpful in assisting participants to understand transformation than textual descriptions  ($\mu =4.45$, $\sigma =1.43$): $U = 34.5$, $p < 0.01$.
In terms of interpretability, 
our technique ($\mu =6.05$, $\sigma =0.95$) was easier to understand than textual descriptions ($\mu =4.40$, $\sigma =1.39$): $U = 71$, $p < 0.01$.

\begin{figure*}[t]
    \centering
    \includegraphics[width=\linewidth]{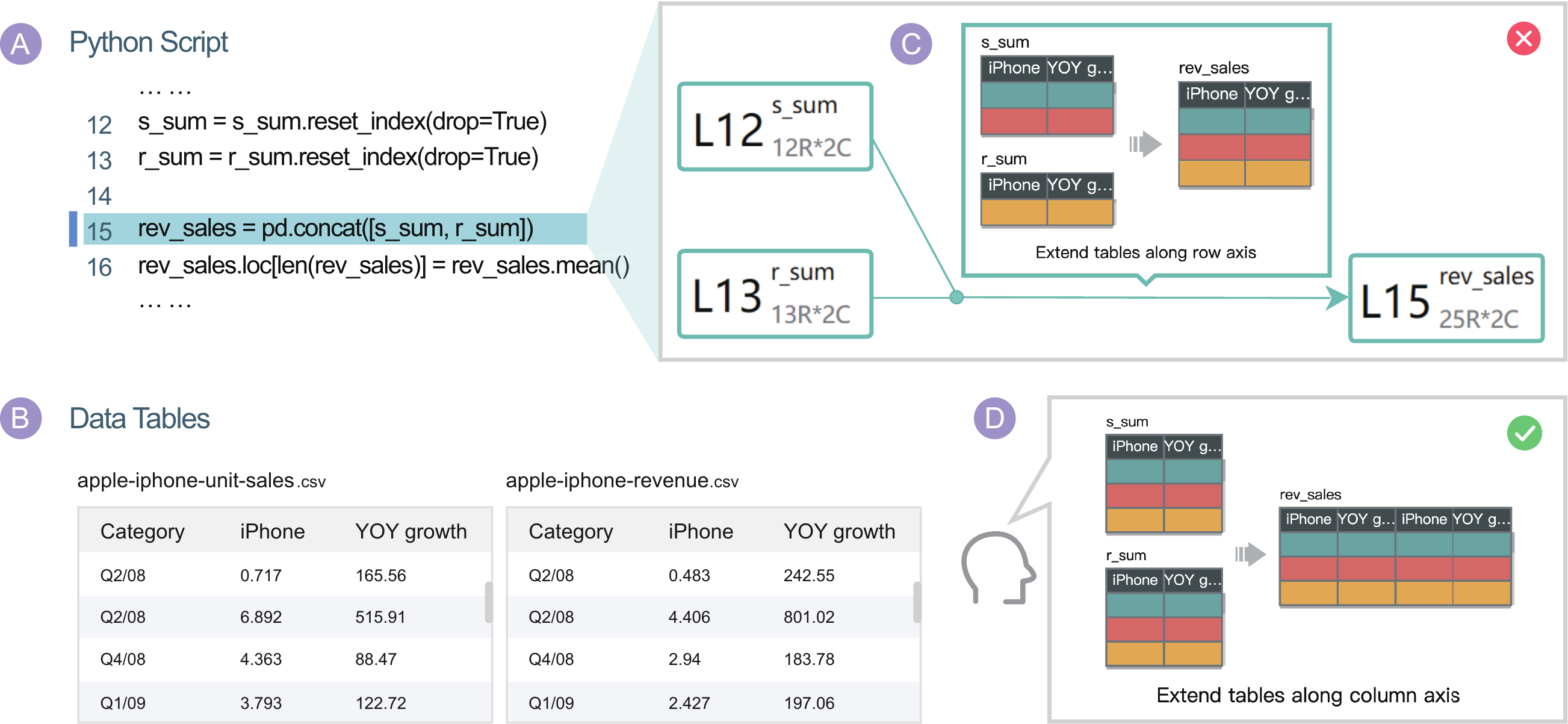}
    \caption{The application of \name{} in validating the process of data transformation written in Python. \name{} takes a piece of code (a) and data tables (b) as input, and outputs a visualization showing table provenance across data transformations. (c) shows a snippet of the visualization. By exploring the table provenance, a data worker can identify errors in the transformation with ease. (d) depicts the correct transformation beared in mind by data workers.}
    \label{fig:teaser}
    \vspace{-0.15cm}  
\end{figure*}

\subsection{\revise{Qualitative Feedback}}

\revise{
All participants showed great interest in the visualization design.
Some participants pointed out that data visualization is a universal language that simplifies learning and communicating. 
For the \textit{Func} tasks, participants appreciate the design of glyphs.
As P6 mentioned that \textit{``it (the glyph visualization) is intuitive and informative.''}
P14 added, \textit{``I do not need to look up the documentation of functions because glyphs explain all.''}
For the \textit{Script} tasks, the node-link diagram presents table provenance using a sequence of glyphs, which is efficient for navigation.
As P3 noted that, \textit{``It is laborious to extract the table dependencies from textual descriptions compared to the visualization.''}
Besides, participants also provided valuable suggestions for our design.}

\revise{
\textbf{Comments on the glyphs:}
The glyph design can be improved from the following aspects.
First, the color encoding of different glyphs may be confusing.}
P8 noted, \textit{``I would think they (columns with the same color in different glyphs) are the same columns.''}
She further suggested, \textit{``Different columns should be depicted using a different color (in the provenance graph).''}
\revise{Second}, some participants (P10, P14) pointed out that text was superior to visualization in some cases. 
For example, the \textit{filter} function in R deletes rows due to some conditions. 
When multiple conditions are passed as parameters, our glyph design cannot distinguish whether BOTH conditions are applied, or EITHER condition is used.
On the contrary, a textual description can articulate it clearly.
P18 and P7 recommended integrating textual description and visualization to utilize the strength of the two techniques.

\revise{
\textbf{Comments on the provenance graph:}
The design of the provenance graph may suffer from two issues.}
First, the provenance graph would be too long when the number of transformations increases. 
As a result, participants continuously zoomed and panned the graph in finishing tasks.
P2 suggested that the pipeline could be presented vertically so that he could explore it using a scroll bar.
In addition, P10 commented, \textit{``The pipeline (provenance graph) should be folded by default, and can be expanded on demand.''}
Second, some participants (P3, P5) suggested supporting programs with complex control flow.
P6 commented, \textit{``The statements in the study are too simple. 
I wonder how the visualization performs in programs with IF-ELSE (conditional statement) or FOR (loop statement) statements.''}

\begin{figure*}[t]
    \centering
    \includegraphics[width=\textwidth]{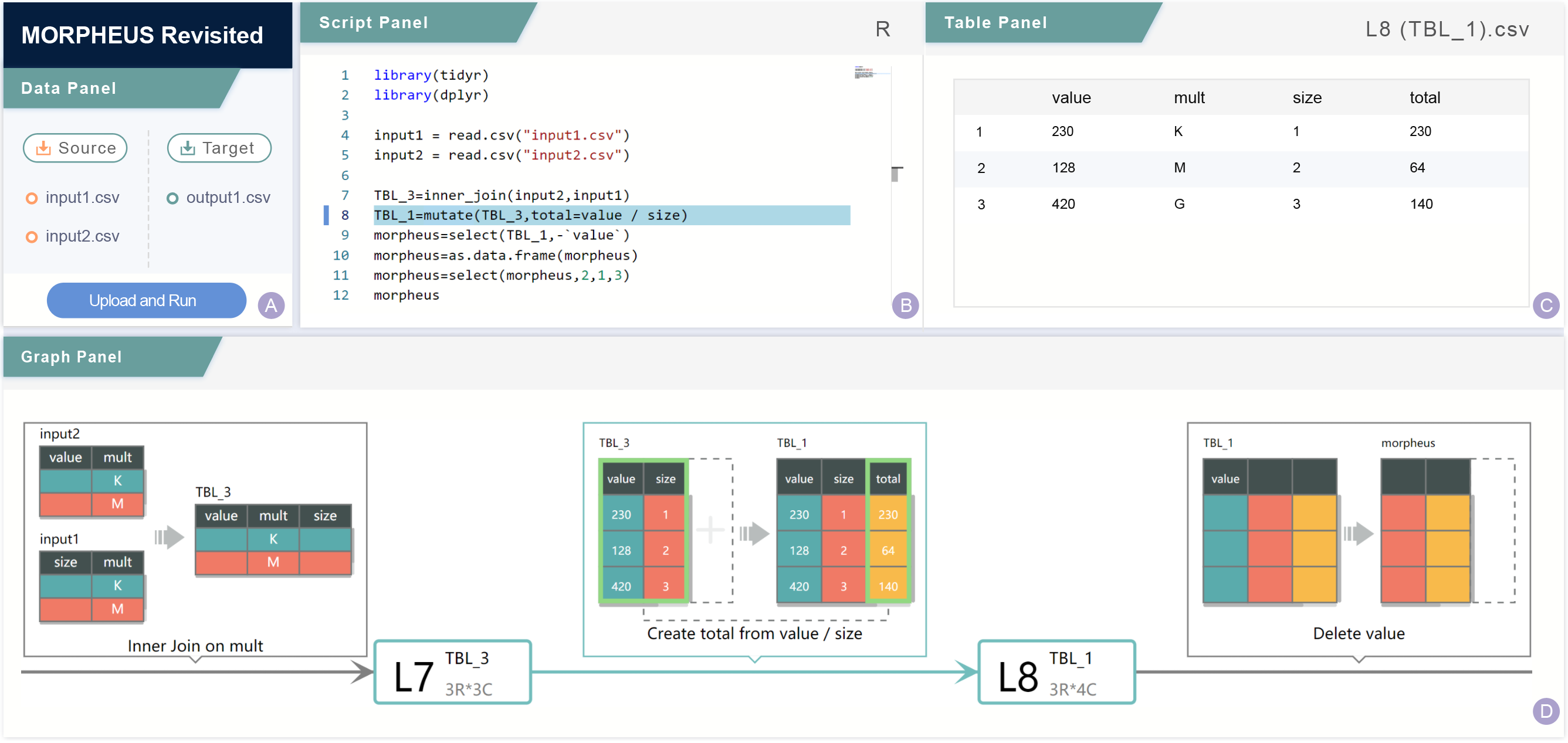}
    \caption{By combining {\scshape Morpheus}, our system generates and illustrates a series of data transformations given source data tables and a target table. The system comprises four panels, \ie, a Data Panel allowing users to upload data tables, a Script Panel showing code pieces in R, a Table Panel shows intermediate data tables, and a Graph Panel that depicts table provenance.}
    \label{fig:reverse_engineering}
    \vspace{-0.08cm}  
\end{figure*}

\section{Example Applications}
To demonstrate how \name{} can be applied to different usage scenarios, we design and implement two prototypes based on \name.
The first prototype helps data scientists validate the procedure of data transformation written in Python, while the second reveals intermediate data transformations given source and target tables. 

\subsection{Double-checking}
\label{subsec:doublecheck}
In this study, we collaborate with two data scientists in a national research lab. 
They usually work together to finish an analytical report.
One critical task in their work is validation, or called double-checking in practice.
Specifically, when a data scientist finishes a workflow, the other needs to check and validate the entire workflow by scrutinizing the code and independently reproducing the workflow.
However, identifying errors in the code is not an easy task, which requires a deep understanding of a large number of functions and data models. 
Inspired by the real-world use case, \kai{the first application illustrates how \name{} aids data scientists in validating and debugging a script of the wrangling process.}


Assume Lucy and Jane are two data scientists working in a national research lab. 
After Jane finishes a data transformation procedure written in Python, Lucy is invited to validate the piece of code to ensure accuracy.
The goal of the code is to combine two tables~\cite{Bare2021Apple} and compute average iPhone unit scales and revenue across years.
Lucy first uploads the two input tables (Figure~\ref{fig:teaser}(b)) in the Data Panel and copy-pastes a code piece to the Script Panel.
After clicking ``Upload and Run'', the provenance graph is displayed in the Graph Panel.
To examine each step, Lucy explores individual transformations in the provenance graph by zooming and panning.
A glyph showing the combination of two tables catches her eye.
As shown in Figure~\ref{fig:teaser}(c), the two tables are combined along the row axis.
However, Lucy makes sure that the two tables should be concatenated by column  (Figure~\ref{fig:teaser}(d)).
To reason the result, she clicks the glyph to locate and highlight the $15$th line of code (Figure~\ref{fig:teaser}(a)).
She notices that the parameter \textit{axis} is not explicitly mentioned in the \textit{concat} function.
By default, however, the concatenation is performed along the row axis with implicit \textit{axis=0}.
Hence, Lucy corrects the statement to \textit{rev\_sales=pd.concat([s\_sum,r\_sum], axis=1)}, and finally obtains the correct results.

\subsection{{\scshape Morpheus} Revisited}

{\scshape Morpheus}~\cite{feng2017component} is a program synthesis algorithm that generates a script for data processing. 
The algorithm accepts multiple source tables and a target table as input and \kai{automatically outputs lines of R code} to reflect the process of transformation.
{\scshape Morpheus} is useful in a number of scenarios.
For example, the output script can automate the process of data transformation and can be reused and revised for future applications.
The output of {\scshape Morpheus}, however, is hard to understand due to obscure function usage and parameters.
In the second application, we apply \name{} to explain the scripts generated by {\scshape Morpheus}.
\revise{The adapted system shown in Figure~\ref{fig:reverse_engineering} is almost identical to the \name{} system.}
The difference is threefold.
First, the Data Panel accepts multiple source tables and a target table, which are passed to {\scshape Morpheus} \revise{on the server side}.
Second, based on code pieces returned by {\scshape Morpheus} and data tables, \name{} runs and \revise{yields} a series of input and output tables for each function and a table provenance and passes them to the \revise{client side}.
Third, the Script Panel shows the script that is not editable.

The application is motivated by a real-world case from StackOverflow~\cite{stackoverflow2015}. 
Assume Devin has two original tables (\textit{input1.csv, input2.csv}) and one target table (\textit{output1.csv}) at hand. 
He first uploads those tables to the system to understand the correct approach to transforming the original tables to the target table.
After clicking the ``Upload and Run'' button,
a piece of code is shown in the Script Panel, and a provenance graph is displayed in the Graph Panel.
From the provenance graph, Devin sees seven edges, indicating that the entire process takes seven data transformations.

Devin has no prior knowledge about R and dplyr.
To get an idea of individual functions, such as \textit{mutate}, he clicks the $8$th line of code in the Script Panel (Figure~\ref{fig:reverse_engineering}(b)). 
Then the transformation and its input and output tables are located and highlighted in the Graph Panel (Figure~\ref{fig:reverse_engineering}(d)).
He figures out that the function creates a new column called \textit{``total''} from \textit{``value''} divided by \textit{``size''}.
From the Script Panel, Devin observes two \textit{select} functions.
He clicks the two lines of code and finds they perform different transformations, \ie, one removes a column \textit{``value''} while the other rearrange columns.

%% file: src/discussion.tex
\section{Discussion}
\label{sec:discussion}

The evaluation shows that our glyph collection and \name{} are effective in presenting data transformations, and \name{} can be generalized to various programming languages and example applications.
Besides feedbacks and suggestions listed in Section~\ref{sec:userstudy}, we identify some limitations in the design and implementation of \name{}.

First, the scalability of \name{} is limited in terms of the number of functions and parameter combinations.
The Code Parser and Transformation Inference modules of \name{} are customized for each function and parameter.
\kai{The current prototype supports $25$ functions from tidyr and dplyr in R and ten functions from Pandas in Python with a small set of parameters. 
Extending our work to a number of functions and parameters is possible. 
However, \revise{it would be tedious.}}
To align with a large number of functions that are typically used for data transformation, a promising direction is to explore learning-based algorithms that can map a function and its parameters to a type of transformation at scale.
This work can act as a starting point for generating training data for such algorithms.
In addition, if a lengthy script contains numerous operations,
\revise{the provenance graph would be too long to navigate.
We acknowledge that the basic layout of transformation workflows will result in node-link diagrams with a suboptimal aspect ratio that require frequent panning/zooming.
We notice that a number of interaction techniques are designed to navigate lengthy content.
For example, focus+context screens~\cite{baudisch2002keeping} can facilitate the exploration of multiscale documents.
In addition, the collapse-to-zoom navigation\cite{baudisch2004collapse} is proposed to explore lengthy web pages. 
In future iterations, we plan to integrate advanced interaction techniques to alleviate the issue.}

Second, the generalizability of the glyph space has yet been explored.
By depicting in-table and out-table text in the glyph, the glyph collection can be generalized to a larger number of transformations, as shown in Figure~\ref{fig:in_table_text}.
On the other hand, the glyph design lacks support for some commonly used data transformations, such as \textit{Transpose}.
To what extent does the glyph space adapts to transformations is unknown.
In future research, we plan to explore the mapping between the glyph space and transformation space to understand the scope.

Third, the presentation of in-table text may result in inconsistencies in some data tables.
For example, in Figure~\ref{fig:operations_glyphs}, the \textit{Combine Rows} shows the results of the \textit{colMeans} function in R, which derives the mean value for each column.
However, the mean value of the input table does match the results in the output table.
That is because all in-table text is derived from the original data. 
In the current stage, we combine textual description and visualization in the glyph design to alleviate the weakness caused by inconsistencies.
\revise{
In addition, if the text in glyphs, including column names, cell contents, and summary descriptions, is long, it can not be fully displayed by default.
This can be difficult for users to spot their difference, especially when the text has the same prefix.
Currently, the text elision issue is resolved through interaction. That is, when a user hovers over the omitted text in the glyph, the whole text will be displayed.}

Fourth, the \revise{shortcoming} of individual glyphs has yet been explored. 
Though the controlled study reveals the overall effectiveness of visual description compared to textual description (\textit{Func} in Figure~\ref{fig:qualitative_result1}), it is far from enough to exploit the \kai{shortcoming} of individual glyphs, which requires enumerating possible functions and their parameters.
For example, the glyph designed for \textit{Delete Rows} may be ineffective when 
a data table does not contain counterexamples.
Because the semantics of filtering conditions can hardly be visualized without counterexamples.
To obtain a comprehensive understanding of the performance of the glyph collection, we plan to validate individual glyphs using various functions and parameter combinations and conduct a large-scale user study for evaluation.

Fifth, the color encoding of glyphs may cause confusion in the provenance graph. 
As one participant in the user study pointed out, columns with the same color in different glyphs may be mistakenly regarded as identical columns.
We intend to combine visual designs with interactions to resolve this ambiguity in our future work.

\section{Conclusion and Future Work}
In this paper, we develop visualization techniques to illustrate the semantics of code pieces in the context of data transformation.
To present individual transformations, we explore design space consisting of two primary dimensions, \ie, key parameters to be encoded and possible visual channels that can be mapped.
Based on the design space, we derive a collection of glyphs targeting $23$ types of transformations.
We argue that the glyph collection can adapt to a broader range of transformations by depicting in-table and out-table text.
To illustrate a sequence of statements, we design and develop \name{}, a pipeline that accepts code pieces and data tables as input and generates a graph showing table provenance across a series of data transformations.
The results of a controlled study have demonstrated the effectiveness and intuitiveness of the visualization design \revise{for our study participants}.
Through example applications, we show how \name{} can be adapted to different programming languages and usage scenarios.

In the future, we plan to enhance \name{} by supporting a large number of functions and parameters in dplyr (R), tidyr (R), and Pandas (Python).
However, the manual enhancement could be laborious and tedious. 
We plan to investigate algorithms that automatically map statements to data transformations to facilitate the adaption of functions and parameters.
Next, since complex control flow is commonly used in data transformation, we would like to explore how to visualize conditional statements and loops in the provenance graph.